\PassOptionsToPackage{unicode}{hyperref}
\PassOptionsToPackage{hyphens}{url}
\PassOptionsToPackage{dvipsnames,svgnames,x11names}{xcolor}
\documentclass[
  12pt]{article}

\usepackage{amsmath,amssymb}
\usepackage{iftex}
\ifPDFTeX
  \usepackage[T1]{fontenc}
  \usepackage[utf8]{inputenc}
  \usepackage{textcomp} 
\else 
  \usepackage{unicode-math}
  \defaultfontfeatures{Scale=MatchLowercase}
  \defaultfontfeatures[\rmfamily]{Ligatures=TeX,Scale=1}
\fi
\usepackage{lmodern}
\ifPDFTeX\else  
\fi
\IfFileExists{upquote.sty}{\usepackage{upquote}}{}
\IfFileExists{microtype.sty}{
  \usepackage[]{microtype}
  \UseMicrotypeSet[protrusion]{basicmath} 
}{}
\makeatletter
\@ifundefined{KOMAClassName}{
  \IfFileExists{parskip.sty}{%
    \usepackage{parskip}
  }{
    \setlength{\parindent}{0pt}
    \setlength{\parskip}{6pt plus 2pt minus 1pt}}
}{
  \KOMAoptions{parskip=half}}
\makeatother
\usepackage{xcolor}
\makeatletter
\ifx\paragraph\undefined\else
  \let\oldparagraph\paragraph
  \renewcommand{\paragraph}{
    \@ifstar
      \xxxParagraphStar
      \xxxParagraphNoStar
  }
  \newcommand{\xxxParagraphStar}[1]{\oldparagraph*{#1}\mbox{}}
  \newcommand{\xxxParagraphNoStar}[1]{\oldparagraph{#1}\mbox{}}
\fi
\ifx\subparagraph\undefined\else
  \let\oldsubparagraph\subparagraph
  \renewcommand{\subparagraph}{
    \@ifstar
      \xxxSubParagraphStar
      \xxxSubParagraphNoStar
  }
  \newcommand{\xxxSubParagraphStar}[1]{\oldsubparagraph*{#1}\mbox{}}
  \newcommand{\xxxSubParagraphNoStar}[1]{\oldsubparagraph{#1}\mbox{}}
\fi
\makeatother

\usepackage{longtable,booktabs,array}
\usepackage{calc} 
\usepackage{etoolbox}
\makeatletter
\patchcmd\longtable{\par}{\if@noskipsec\mbox{}\fi\par}{}{}
\makeatother
\IfFileExists{footnotehyper.sty}{\usepackage{footnotehyper}}{\usepackage{footnote}}
\makesavenoteenv{longtable}
\usepackage{graphicx}
\makeatletter
\def\maxwidth{\ifdim\Gin@nat@width>\linewidth\linewidth\else\Gin@nat@width\fi}
\def\maxheight{\ifdim\Gin@nat@height>\textheight\textheight\else\Gin@nat@height\fi}
\makeatother
\setkeys{Gin}{width=\maxwidth,height=\maxheight,keepaspectratio}
\makeatletter
\def\fps@figure{htbp}
\makeatother

\makeatletter
\@ifpackageloaded{caption}{}{\usepackage{caption}}
\AtBeginDocument{%
\ifdefined\contentsname
  \renewcommand*\contentsname{Table of contents}
\else
  \newcommand\contentsname{Table of contents}
\fi
\ifdefined\listfigurename
  \renewcommand*\listfigurename{List of Figures}
\else
  \newcommand\listfigurename{List of Figures}
\fi
\ifdefined\listtablename
  \renewcommand*\listtablename{List of Tables}
\else
  \newcommand\listtablename{List of Tables}
\fi
\ifdefined\figurename
  \renewcommand*\figurename{Figure}
\else
  \newcommand\figurename{Figure}
\fi
\ifdefined\tablename
  \renewcommand*\tablename{Table}
\else
  \newcommand\tablename{Table}
\fi
}
\@ifpackageloaded{float}{}{\usepackage{float}}
\floatstyle{ruled}
\@ifundefined{c@chapter}{\newfloat{codelisting}{h}{lop}}{\newfloat{codelisting}{h}{lop}[chapter]}
\floatname{codelisting}{Listing}

\makeatother
\makeatletter
\@ifpackageloaded{caption}{}{\usepackage{caption}}
\@ifpackageloaded{subcaption}{}{\usepackage{subcaption}}
\makeatother

\ifLuaTeX
  \usepackage{selnolig}  
\fi
\usepackage[]{natbib}
\usepackage{bookmark}

\IfFileExists{xurl.sty}{\usepackage{xurl}}{} 
\hypersetup{
  pdftitle={Title},
  pdfauthor={Author 1; Author 2},
  pdfkeywords={3 to 6 keywords, that do not appear in the title},
  colorlinks=true,
  linkcolor={blue},
  filecolor={Maroon},
  citecolor={Blue},
  urlcolor={Blue},
  pdfcreator={LaTeX via pandoc}}

\usepackage{amsthm}

\def\spacingset#1{\renewcommand{\baselinestretch}%
{#1}\small\normalsize} \spacingset{1}

\newtheorem{example}{Example}
\newtheorem{theorem}{Theorem}[section]
\newtheorem{lemma}[theorem]{Lemma}
\newtheorem{proposition}[theorem]{Proposition}
\newtheorem{corollary}[theorem]{Corollary}
\newtheorem{assumption}[theorem]{Assumption}
\newtheorem{definition}[theorem]{Definition}
\newtheorem{remark}[theorem]{Remark}

\usepackage{algorithm}
\usepackage{algorithmic}

\usepackage{listings}
\usepackage{tikz}
\usepackage{subcaption}
\usepackage{lineno}
\newcommand{\anon}{1}
\begin{document}

\bibliographystyle{agsm}

\spacingset{1}


\if0\anon
{
  \title{\bf A General Marked Point Process Framework For Self-Exciting Network Evolution}
  \maketitle
} \fi

\if1\anon
{
  \title{\bf A General Marked Point Process Framework For Self-Exciting Network Evolution}
  \author{Duncan A. Clark \hspace{.2cm}\\
    Department of Statistics, Williams College\\
    and \\
    Conor J. Kresin \\
    Department of Mathematics and Statistics, University of Otago\\
    and \\
    Charlotte M. Jones-Todd\\
    Department of Statistics, University of Auckland}
  \maketitle
} \fi

\begin{abstract}
We propose a novel modeling framework for time-evolving networks allowing for long-term dependence in network features that update in continuous time. Dynamic network growth is functionally parameterized via the conditional intensity of a marked point process. This characterization enables flexible, joint modeling of both update timing and the network updates themselves, dependent on the entire left-continuous sample path. We propose a path dependent nonlinear marked Hawkes process as an expressive platform for modeling such data; its dynamic mark space embeds the time-evolving network. We prove well-posedness and establish sufficient stability conditions, demonstrate simulation and subsequent feasible likelihood-based inference through numerical study, and illustrate the methodology with an application to conference attendee social network data. The proposed formulation provides a flexible and principled foundation for statistical inference on complex network evolution in continuous time.
\end{abstract}

\noindent%
\vfill
\newpage
\spacingset{1.8} 

\section{Introduction}

Dynamic networks are well suited to modeling a variety of applications ranging from protein interaction \citep{jeong2001} to co-authorship \citep{newman2001}, economic behavior \citep{jackson2002evolution}, public health state laws \citep{clark2024}, and friendship networks \citep{lakon2014}. Such applications are characterized by a set of edges and nodes whose evolution exhibits complex temporal and network dependencies. There are many approaches to understanding the global properties of such networks, from early Erd\H{o}s--R\'enyi models \citep{erdos59}, to small-world models \citep{watts1998}, to scale-free approaches \citep{BA_model,BA_fitness_2001}, and densification over time \citep{Leskovec2005}. There are also approaches that seek to model bursty temporal networks; examples are found in \citet{sheng_bursty,holme_2012,holme_2015} and references therein. Such methods are useful for reproducing quantitative structure (centrality, path lengths, degree distributions, clustering), but they typically do not provide likelihood-based statistical inference for the network evolution mechanism.

Inference-focused dynamic network models often regard the node set as fixed and target the edge formation process over time. Examples include Stochastic Actor Oriented Models (SAOM) \citep{snijdersSAOM}, Temporal Exponential-family Random Graph models (TERGM) \citep{hanneke2010discrete}, Separable TERGM (STERGM) \citep{krivitsky_STERGM}, latent space Bayesian approaches \citep{Hoff2015}, and Latent Order Logistic models (LOLOG) \citep{Fellows2018}. These methods provide statistical inference on the edge formation process, but the node set is treated as fixed, so they do not facilitate inference on node arrivals.

However, in many settings, the network update times and the nature of those updates are not credibly independent \citep{snijders_2001}. Self-excitation in time is prevalent \citep{HuangLatentHawkes2022}: the arrival of a node or edge can increase the propensity for further updates in the near future. At the same time, network topology shapes which updates are plausible: preferential attachment, triadic closure, and other local mechanisms depend on the current network state. Capturing this joint dependence requires a continuous-time model in which update times and update content are jointly modeled.

Therefore, we propose a novel marked point process  \citep{daley2003introduction,kallenberg1976random} characterization of time evolving networks: each network update corresponds to an event time, and the mark attached to that time is the update itself (the new nodes and edges added). This perspective allows a single conditional intensity characterization for the coupled evolution of timing and topology. We focus on a Hawkesian ground specification \citep{Hawkes1971,hawkes1974cluster}, since self-excitation is a natural way to represent bursts of activity. However, the mark-space embedding that we propose is compatible with other point-process specifications. Our marked point process viewpoint enables hitherto impossible statistical inference in this growing network setting, affording interpretability via parametric specification.

Previous work connects point process and network modeling paradigms by treating node interaction data as realizations of continuous-time stochastic processes. For example, early relational event models \citep{Butts2008, VuHunterSmythAsuncion2011, HunterSmythVuAsuncion2011} and multivariate counting representations \citep{PerryWolfe2013} represent dynamic interactions  as collections of multivariate point processes indexed by dyads. However, these approaches do not model endogenous excitation through self-exciting intensity kernels. The temporal dependence is simply introduced via history statistics or time-varying covariates.

\citet{PassinoHeard2023} introduced mutually exciting point process graphs, where Hawkes-type intensities model the proposed self-exciting interaction dynamics on a fixed set of nodes, with intensities defined over dyads. Here, latent node parameters induce edge-specific baseline intensities and mutual excitation effects across dyads. Similarly \citet{HuangLatentHawkes2022} proposed a latent space Hawkes formulation modeling how network structure and temporal excitation jointly influence node interactions through evolving latent positions, rather than through explicit stochastic network updates. The recent generative framework of \citet{Perez2025} attempts to characterize temporal networks, however they rely on a restrictive first-order Markovian assumption where the intensity is reduced to a function of the most recent event and a vector of hand-crafted features. In contrast, our framework avoids these simplifying losses of information, capturing the deep, long-term dependencies inherent in evolving network structures.

The closest precedent to our proposed framework is \citet{farajtabar2017coevolve}, where network updates are modeled via a joint conditional intensity and survival function in a multivariate Hawkes specification. While such work accommodates growth, inference on the network update mechanism is limited. The path-wise connection between univariate marked Hawkes processes and multivariate representations \citep{davis2024multivariate} provides a bridge to a large literature on multivariate Hawkes processes in network contexts, including mean-field limits \citep{delattre2016hawkes}, Hawkes graphs \citep{embrechts2018hawkes}, and latent group structure \citep{fang2024group}. We adopt a univariate marked formulation for parsimony: marks encode updates directly, avoiding the explosion of dimension that arises when treating each potential interaction type as a separate component.

The paper is structured as follows. Section~\ref{sec:prelim} introduces notation and the dynamic mark space encoding network updates. Section~\ref{sec:induced_models} presents our proposed ``HawkesNet'' models induced by network update probability mass functions (PMFs) and develops likelihood-based inference. Section~\ref{sec:path_dependent} introduces mark path dependent Hawkes processes and establishes stability results. Section \ref{sec:hypertext} describes an application to dynamic human contact patterns at the 2009 ACM Hypertext conference. All results in this paper were derived using the publicly available \texttt{hawkesNet} software package 
\if0\anon (citation blinded for peer review) \else \cite{hawkesNet}.
\fi

\section{Dynamic Network Point Process Setup}\label{sec:prelim}

We consider a network at time $t$ to have  $N_t$ nodes, with edges represented by a binary matrix $A_t = \lbrace A_{i,j}^t \rbrace_{i=1,j=1}^{N_t}$ with $A^{t}_{i,j} \in \lbrace 0,1\rbrace$. The element $A_{i,j}^t$ is 0 if node $i$ and node $j$ do not have an edge and $1$ if they do,  at time $t$. The network is defined as $\mathcal{G}_t = \lbrace N_t,A_{t} \rbrace $ at time $t$. Each edge and node in $\mathcal{G}_t$ has an associated birth, $b_i$ and $b_{i,j}$ respectively, defined as the earliest time it appears in the graph. Define 
\begin{align*}
    b_i =:& \inf(\lbrace t : i \leq N_t\rbrace)\\
    b_{i,j} =:& \inf(\lbrace t : A_{i,j}^{t} = 1\rbrace ).
\end{align*} 
At time $T$ we can consider the node and edge arrival times $\mathcal{T}_{T} = \lbrace b_{j} : j\in \lbrace 1\ldots N_T \rbrace \rbrace \cup \lbrace b_{j,k} : j,k \in \lbrace 1\ldots N_T \rbrace \rbrace$.

 The node arrivals $b_i$ and edge arrivals $b_{i,j}$  may overlap, denoting an edge and a node being added to the network at the same time. Thus, we let $t_i \in \mathcal{T}_{T}$ be the arrival time of the $i$th network update. This is now a marked point process; see Section \ref{sec:mpp_notation} for precise specification. Each event is an update to the network, which allows both edges and nodes to be added at the same time, see Section \ref{sec:mpp_updates} for formal mark definitions.

\subsection{Marked point process notation}\label{sec:mpp_notation}

Following \citet{daley2007introduction}, a marked point process (MPP) is a $\mathbb{Z}^+$-valued random measure on a complete separable metric space (CSMS) $\mathcal X$, in this work we set $\mathcal X=\mathbb{R}^+\times\mathcal{M}$, where $\mathcal{M}$ denotes the mark space. The MPP, $N$, has associated ground process $N_g(\cdot)$ which corresponds to temporal locations of points: $N_g(A)=N(A\times\mathcal M)$. We assume $N$ is locally finite and simple. On a bounded time window $[0,T)$ a realization consists of a finite set of tuples $\{(t_i,m_i)\}_{i=1}^{N([0,T)\times \mathcal M)}$. Let $\mathcal H_t$ be the natural history up to but not including time $t$, i.e.\ $\mathcal H_t=\sigma(N([0,t)\times B):B\subseteq \mathcal M)$. The conditional intensity of $N$, denoted $\lambda$, is an integrable, nonnegative, $\mathcal{H}_t$-predictable process such that
\[
\mathbb{E}[N(dt,dm)\mid\mathcal H_t]=\lambda(t,m\mid\mathcal H_t)\,dt\,\nu(dm),
\]
where $\nu$ is a reference measure on $\mathcal M$. Since $\mathcal M$ will be discrete in our construction, the natural choice is counting measure.

 A point process is simple if, with probability one, all the points' spatiotemporal locations are distinct. We consider only simple point process estimation. Since the conditional intensity, $\lambda$, uniquely determines the finite-dimensional distributions of any simple stationary point process (Proposition 7.2.IV of \cite{daley2003introduction}) we can model an MPP by specifying a model for $\lambda$. We note that growing networks themselves are often non-stationary; conveniently, the MPP characterization itself is stable (see Section \ref{sec:path_dependent} for further discussion). 

Self-exciting point processes, where points in the past affect future intensity, can be modeled by Hawkes processes \citep{Hawkes1971}. A marked Hawkes process with background rate $\lambda_0(m)$ and triggering kernel $g$ is characterized in Definition \ref{def:mpp_hawkes}.

\begin{definition}[Marked Hawkes process]\label{def:mpp_hawkes}
Let $N$ be a simple marked point process on $\mathbb{R}_+ \times \mathcal{M}$ with
natural filtration $\{\mathcal{H}_t\}_{t \ge 0}$.  
We say that $N$ is a marked Hawkes process if its
$\mathcal{H}_t$-conditional intensity admits the representation
\begin{equation}
\lambda(t,m \mid \mathcal{H}_t)
=
\lambda_{\emptyset}(t,m)
+
\int_{(0,t)\times\mathcal{M}}
g\!\left(t-u, m,m'\right)\,
N(du,dm'),
\label{eq:marked_hawkes}
\end{equation}
where $\lambda_{\emptyset}(t,m)$ is a predictable baseline intensity and
$g(t,m,m')$ is the excitation kernel.
\end{definition}

Classical (linear) Hawkes processes have well-known non-explosion/stability theory under subcriticality of the triggering kernel \citep{Hawkes1971} or in the marked case, when expected finite cluster size (Lemma 6.3.II of \citealp{daley2007introduction}). For self-exciting network growth process, a network update at time $t$ depends on the current network as well as the timing and nature of previous updates. Such a process is termed non-separable.

Separability is often assumed to facilitate likelihood computation \citep{davis2024multivariate,spassiani2024distribution}, and is not to be confused with the assumption of independent or unpredictable marks \citep{daley2007introduction}. Intuitively, a process is mark separable if the time process (corresponding to the ground process $N_g$, see \cite{daley2003introduction} for more details) and mark distribution can be decoupled after conditioning on history. This restricts the rich class of models which require jointly considering how marks affect future event rates.

A non-separable specification is natural for the time-evolving random network growth process setting. For example, one may expect to wait a short time for a background singleton node to be added to the network, but a relatively long time for a new highly connected node to arrive. For discussion and development of this idea see Section \ref{sec:path_dependent}.

\begin{remark}[Topology--timing feedback vs.\ the ground--mark decomposition]
\label{rem:topology_timing_vs_decomp}
Two notions are useful to keep distinct.

\emph{(i) Ground--mark decomposition (algebraic).}
Whenever the ground intensity is finite, any marked point process admits the identity
\begin{equation}
\label{eq:algebraic_factorization}
\lambda(t,m\mid\mathcal H_t)=\lambda_g(t\mid\mathcal H_t)\,q(m\mid t,\mathcal H_t),
\qquad
\lambda_g(t\mid\mathcal H_t):=\sum_{m\in\mathcal M}\lambda(t,m\mid\mathcal H_t).
\end{equation}
with $q(\cdot\mid t,\mathcal H_t)$ the conditional mark PMF
$q(m\mid t,\mathcal H_t)=\lambda(t,m\mid\mathcal H_t)/\lambda_g(t\mid\mathcal H_t)$ (when $\lambda_g>0$).
We exploit this identity for likelihood evaluation in Section~\ref{sec:likelihood}.

\emph{(ii) Non-separability (substantive).}
In this paper, non-separability refers to topology--timing feedback: the evolving network state influences the rate of updates
and/or the temporal excitation regime, and network updates in turn reshape the topology that governs future timing and future update preferences.
A model can therefore admit the decomposition in (i) and still be non-separable in this substantive sense.
\end{remark}

\subsection{Network evolution as a marked point process}\label{sec:mpp_updates}

The network growth path $(\mathcal G_t)_{t\ge 0}$ is piecewise constant and changes only when new node(s) and/or edge(s) appear, we write $\mathcal G_{t-}$ for the left limit. At each $t_i$ we collect all new objects into a single mark $m_i$.

Let us consider a general space of possible updates to a network. Let $\mathcal V$ be a countably infinite set of possible node labels, and let $\mathcal E$ be the set of all potential edges on $\mathcal V$ (e.g.\ $\mathcal E=\{\{u,v\}:u,v\in\mathcal V,\ u\neq v\}$ for an undirected simple graph). Each event mark is a finite, nonempty subset
\[
m_i \subseteq \mathcal V \cup \mathcal E,
\qquad m_i\neq \emptyset,\quad |m_i|<\infty.
\]
Elements of $m_i\cap\mathcal V$ are the new nodes born at $t_i$, and elements of $m_i\cap\mathcal E$ are the new edges born at $t_i$. Edges in $m_i$ may connect pre-existing nodes and/or nodes introduced simultaneously in $m_i$.

The corresponding mark space is
\[
\mathcal M := \bigl\{m\subseteq \mathcal V\cup\mathcal E:\ m\text{ finite and nonempty}\bigr\}.
\]
Because $\mathcal V$ is countable, so is $\mathcal E$, and hence $\mathcal M$ is countable. While $\mathcal M$ is fixed as a set, the admissible marks evolve with the current network state. Let $\mathcal M_t\subseteq\mathcal M$ denote the set of marks that correspond to valid additions given $\mathcal G_{t-}$ (e.g.\ no duplicate nodes/edges relative to $\mathcal G_{t-}$, and every edge endpoint is either already present at $t-$ or is introduced by the same mark). Admissibility is enforced by requiring $ \lambda(t,m\mid\mathcal H_t)=0$ for $m\notin \mathcal M_t.$

Functionally we can understand this as a dynamic mark space, i.e. the space of possible updates to the network depends on the current network. This is in contrast to typical marked point processes, for example a typical spatiotemporal point process has a fixed $\mathbb{R}^{2}$ mark space.

\begin{definition}[Admissible update marks]\label{def:admissible_marks}
Given the left-limit graph $\mathcal G_{t-}=(V_{t-},E_{t-})$, a mark $m\in\mathcal M$ is admissible at time $t$ if, writing $\Delta V:=m\cap\mathcal V$ and $\Delta E:=m\cap\mathcal E$, the following hold:
\begin{enumerate}
\item $\Delta V \cap V_{t-}=\emptyset$ and $\Delta E \cap E_{t-}=\emptyset$;
\item every edge $e\in \Delta E$ has endpoints in $V_{t-}\cup \Delta V$;
\item $\Delta E$ contains no repeated edges and no self-loops (as appropriate for $\mathcal E$).
\end{enumerate}
We write $\mathcal M_t\subseteq\mathcal M$ for the set of all admissible marks at time $t$.
\end{definition}

This construction makes the key identification used throughout the paper: the evolving network and the marked update sequence are two views of the same object. Given an initial graph $(V_0,E_0)$, define the node and edge-components of a mark by
\[
\Delta V_i := m_i \cap \mathcal V,
\qquad
\Delta E_i := m_i \cap \mathcal E.
\]
Then, given an initial graph $(V_0,E_0)$, the cumulative node and edge sets at time $t$ are obtained by accumulation:
\[
V_t = V_0 \,\cup\, \bigcup_{i:\,t_i\le t}\Delta V_i,
\qquad
E_t = E_0 \,\cup\, \bigcup_{i:\,t_i\le t}\Delta E_i.
\]
Thus $\mathcal G_t$ is determined by $\{(t_i,m_i):t_i\le t\}$. Conversely, the jump times and jump sizes of $(\mathcal G_t)$ define the marked update sequence.

We observe updates $\{(t_i,m_i)\}_{i=1}^n$ on $[0,T)$. We represent them as a simple marked point process
$N=\sum_{i=1}^n \delta_{(t_i,m_i)}$ on $\mathbb R^+\times\mathcal M$, with history $\mathcal H_t$ and conditional intensity
$\lambda(t,m\mid\mathcal H_t)$ w.r.t.\ $dt\,\nu(dm)$ (here $\nu$ is counting measure on the countable $\mathcal M$).
The simplicity of the process is ensured by regarding network updates as marks, not simply restricting to adding single edges or nodes as events.

\section{Network probability mass function induced models}\label{sec:induced_models}

Specifying which network updates are plausible is crucial to modeling network evolution. We consider induced specifications that build a marked intensity from (i) a ground-rate model for when updates occur, and (ii) a network update PMF for which update occurs using the following:
\begin{equation}\label{eq:lambda_factor_induced}
\lambda(t,m\mid\mathcal H_t)=\lambda_g(t\mid\mathcal H_t)\,q(m\mid t,\mathcal H_t),
\end{equation}
where $q(\cdot\mid t,\mathcal H_t)$ is a PMF supported on the admissible set $\mathcal M_t$.
(Recall from Remark~\ref{rem:topology_timing_vs_decomp} that \eqref{eq:lambda_factor_induced} is always valid as an identity; in the induced HawkesNet class we specify $\lambda_g$ and $q$ directly, which avoids an explicit sum over the dynamic mark space in likelihood calculations; see Section~\ref{sec:likelihood}.) We call any model built by combining a ground-rate specification for $\lambda_g$ with a network-growth PMF $q$ a HawkesNet model. Non-separability in this setting allows $\lambda_g$ and $q$ depend on the evolving graph (equivalently, on the full marked history), enabling timing--network topology feedback.

As one of two considered induced models, we formulate a two-parameter $(\tau, m)$ BA-style kernel with time decay based on the Barabasi-Albert (BA) preferential attachment model \citep{BA_model}, which is widely used to explain power-law degree distributions. In this formulation, $\tau > 0$ dictates the decay rate of past degrees, while $m > 0$ represents the expected number of edges added per event.

\begin{example}[Preferential Attachment HawkesNet]\label{ex:BA}

At event time $t$, the mark $m_t$ is drawn as follows: (i) add one new node at time $t$; (ii) draw the number of edges to add, $K_t$, from a distribution with mean $m$ (e.g., $K_t \sim \text{Poisson}(m)$), and set $K_t \leftarrow \min(K_t, N_{t-})$ so that at most $N_{t-}$ edges are added; (iii) connect the new node to $K_t$ distinct existing nodes by sampling without replacement according to the BA attachment probabilities below.

The probability that an existing node $i$ is chosen as an attachment target (given the history $\mathcal{H}_t$) is$ p_i^{\text{BA}} = \frac{\delta_i}{\sum_{k=1}^{N_{t-}} \delta_k}$ where $\delta_i^t = \exp(-\tau (t - t_i)) \cdot d_i^t$, and $d_i^t$ is the degree of node $i$ immediately before time $t$. This yields preferential-attachment behavior with time decay, and $\delta_i^t$ plays the role of a time-decayed ``fitness'' in the extended BA framework \citep{BA_fitness_2001}. Let $\mathcal{E}(m_t)$ denote the edge set of the mark $m_t$, and define $e_i = \mathbb{I}\bigl((N_{t-}+1, i) \in \mathcal{E}(m_t)\bigr)$. The mark density can be written as a product of the marginal distribution for $K_t$ (e.g., $\text{Poisson}(K_t; m)$) and that of sampling $m$ edges without replacement. As $N_{t-}$ is typically large, and edges cannot be selected more than once, for practical purposes we approximate this with a product of Bernoulli terms
\begin{align}\label{eq:bernoulli}
q(m_t \mid t, \mathcal{H}_t) = \mathbb{P}(K_t \mid m) \times \prod_{i=1}^{N_{t-}} \bigl(p_i^{\text{BA}}\bigr)^{e_i} \bigl(1 - p_i^{\text{BA}}\bigr)^{1-e_i}.
\end{align}
\end{example}

The BA model adds nodes to well connected nodes, independently of other structure in the network. For social networks, where existing, complex network structure may impact the probability of an edge being added, this is insufficient. We consider the change statistic model as an alternative, accommodating complex network structure.

\begin{example}[Change Statistic HawkesNet]\label{ex:cs}
For a network $y$ and a vector of network statistics $g(y) \in \mathbb{R}^{p}$, the change statistic for an update $m$ is $C(m) = g(y^{m+}) - g(y^{m-})$, where $y^{m+}$ is the network $y$ with the update $m$ applied and $y^{m-}$ is the network without that update. Change statistics measure the effect of an update on specified network statistics. A mark PMF can be induced by adding edges according to a logistic model on change statistics, following the Exponential-family Random Graph Model (ERGM) literature \citep{FrankStrauss1986,snijders2006,Robins2007, Fellows2018}.
For a single candidate edge $(i,j)$, write $C_{i,j} = g(y_{i,j}^{+}) - g(y_{i,j}^{-})$. For example, if $g(y) = (\text{edge count}(y), \text{triangle count}(y))^{\top}$, then $C_{i,j}$ has first component $1$ and second component equal to the number of triangles that edge $(i,j)$ would complete.
We draw the mark $m_t$ at event time $t$ using parameters $\theta \in \mathbb{R}^{p}$ for the change statistics, $\tau > 0$ for time-decay, $\lambda > 0$ for node arrivals, and $m > 0$ for edge arrivals.
\begin{enumerate}
\item \textbf{Node Arrival:} Draw the number of new nodes, $N_t^{\text{new}}$, from $\text{Poisson}(\lambda)$ and add these nodes at time $t$.
\item \textbf{Edge Count:} Draw the number of edges to add, $K_t$, from $\text{Poisson}(m)$ and restrict $K_t \leq |\mathcal{C}_t|$, where $\mathcal{C}_t$ is the set of candidate edges between new and existing nodes.
\item \textbf{Edge Selection:} For each candidate $(i,j) \in \mathcal{C}_{t}$, define the selection weight:
\begin{equation}\label{eqn:csdraw}
w_{i,j} = \exp\bigl(-\tau \cdot (t - t_i)\bigr) \cdot \frac{1}{1 + \exp(-\theta^{\top} C_{i,j})},
\end{equation}
where $t_i$ is the time of the most recent activity of node $i$. Sample $K_t$ distinct edges from $\mathcal{C}_{t}$ without replacement, with probabilities proportional to $w_{i,j}$.
\end{enumerate}
Thus $m$ governs the density of edges per event, while $\theta$ and $\tau$ govern the structural and temporal preferences of those edges. Under the approximation that edge indicators are conditionally independent given the history, the mark density factors as the product of Poisson masses for nodes and edges, and approximately a product of Bernoulli terms over the candidates as in Example \ref{ex:BA}
\begin{equation}\label{eq:csdensity}
q(m_t \mid t, \mathcal{H}_{t}) = \text{Poisson}(N_t^{\text{new}}; \lambda) \times \text{Poisson}(K_t; m) \times \prod_{(i,j) \in \mathcal{C}_{t}} \bigl(p_{i,j}^{\text{CS}}\bigr)^{e_{i,j}} \bigl(1 - p_{i,j}^{\text{CS}}\bigr)^{1 - e_{i,j}},
\end{equation}
where $e_{i,j} = \mathbb{I}\bigl((i,j) \in \mathcal{E}(m_t)\bigr)$ and $p_{i,j}^{\text{CS}}$ is the normalized weight $w_{i,j}$. $\mathcal{C}_t$ grows with $n^{2}$ so the product of Bernoulli approximation is appropriate for even modestly sized networks.
\end{example}

In Examples~\ref{ex:BA}--\ref{ex:cs} the edge component of the mark is sampled without replacement (from existing nodes in Example~\ref{ex:BA} and from the candidate set $\mathcal C_t$ in Example~\ref{ex:cs}). To minimize notation, we approximate the exact conditional PMF using the conditionally independent product forms in \eqref{eq:bernoulli} and \eqref{eq:csdensity}. The broader HawkesNet framework does not depend on this approximation; exact without-replacement PMFs may be substituted when computationally convenient.

One might think that an intercept term in the change statistic formula would be sufficient to model the baseline propensity of edges. For practical data, this is not the case. Mark updates are typically local, meaning a low number of new nodes or edges is added at each event; thus, having parameters $\lambda$ and $m$ that do not change with time is crucial to fitting realistic data.

There is significant work on similar models \citep{Fellows2018,lusher2012} with social theory driving the selection of suitable statistics \citep{snijders2006}. The model is specified by the choice of network statistics, common choices include triangles for modeling transitive closure, as well as k-stars, (a single node connected to $k$ others) to model node popularity. One can also consider nodal covariate  based statistics, for example gender, one could then consider the count of matched gender edges, or other related statistics. The inclusion of structural and nodal covariate statistics allows for inference on the nodal covariate, while also accounting for the tendency towards transitive closure.

\section{Mark path dependent Hawkes processes}\label{sec:path_dependent}

Linear Hawkes processes admit a Poisson cluster representation \citep{hawkes1974cluster} in which each event contributes a fixed offspring mechanism to future intensity. For network growth this is often too rigid: the effect of an old update can change as the graph evolves. An edge created early may later participate in many triangles or become embedded in a high-degree region of the network, so its influence at time \(t\) depends on the current graph, not only on the mark at its birth. Figures~\ref{fig:networkHawkes} and~\ref{fig:hawkes_linear_vs_nonlinear} illustrate this distinction.

\begin{figure}[!htb]
    \centering
    \begin{tikzpicture}[>=stealth,thick,scale=0.6]

\draw[->] (0,0) -- (0,3.8) node[above] {\(\lambda(t,m\mid\mathcal{H}_t)\)};
\draw[->] (0,0) -- (9,0)   node[right] {\(t\)};

\def\tOne{2.1}
\def\tTwo{4.8}
\def\tThree{6.2}
\def\baseline{1}

\draw[thick] (0,\baseline) -- (\tOne,\baseline);
\draw[thick] (\tOne,\baseline) -- (\tOne,3);
\draw[thick,domain=\tOne:\tTwo,samples=60,smooth]
  plot(\x,{ (3 - \baseline)*exp(-( \x - \tOne)) + \baseline });
\pgfmathsetmacro{\valAtTTwo}{
  (3 - \baseline)*exp(-( \tTwo - \tOne)) + \baseline
}
\draw[thick] (\tTwo,\valAtTTwo) -- (\tTwo,2.7);
\draw[thick,domain=\tTwo:\tThree,samples=60,smooth]
  plot(\x,{ (2.7 - \baseline)*exp(-( \x - \tTwo)) + \baseline });
\pgfmathsetmacro{\valAtTThree}{
  (2.7 - \baseline)*exp(-( \tThree - \tTwo)) + \baseline
}
\draw[thick] (\tThree,\valAtTThree) -- (\tThree,3.2);
\draw[thick,domain=\tThree:9,samples=60,smooth]
  plot(\x,{ (3.2 - \baseline)*exp(-( \x - \tThree)) + \baseline });

\node at (\tOne,0) {\Large $\times$};
\node at (\tTwo,0) {\Large $\times$};
\node at (\tThree,0) {\Large $\times$};

\draw[->] (\tOne,0)   -- (\tOne,-1);
\draw[->] (\tTwo,0)   -- (\tTwo,-1);
\draw[->] (\tThree,0) -- (\tThree+1,-1);

\draw (\tOne,-2) circle (.5);
\node[circle,fill=red,scale=0.3] (a1) at (\tOne-0.2,-2.2) {};
\node[circle,fill=red,scale=0.3] (a2) at (\tOne+0.0,-1.7) {};
\node[circle,fill=red,scale=0.3] (a3) at (\tOne+0.2,-2.2) {};
\draw[red] (a1) -- (a2) -- (a3) -- (a1);

\draw (\tTwo,-2) circle (.75);
\node[circle,fill=black,scale=0.3] (b1) at (\tTwo-0.2,-2.2) {};
\node[circle,fill=black,scale=0.3] (b2) at (\tTwo+0.0,-1.7) {};
\node[circle,fill=black,scale=0.3] (b3) at (\tTwo+0.2,-2.2) {};
\draw (b1) -- (b2) -- (b3) -- (b1);
\node[circle,fill=red,scale=0.3] (b4) at (\tTwo,-2.6) {};
\draw[red] (b2) -- (b4);

\draw (\tThree+1.6,-2) circle (1);
\node[circle,fill=black,scale=0.3] (c1) at (\tThree+1.3,-2.3) {};
\node[circle,fill=black,scale=0.3] (c2) at (\tThree+1.6,-1.7) {};
\node[circle,fill=black,scale=0.3] (c3) at (\tThree+1.9,-2.3) {};
\node[circle,fill=black,scale=0.3] (c4) at (\tThree+1.6,-2.6) {};
\draw (c1) -- (c2) -- (c3) -- (c1) -- (c4) -- (c2);
\node[circle,fill=red,scale=0.3] (c5) at (\tThree+1.2,-1.9) {};
\draw[red] (c4) -- (c5);
\draw[red] (c2) -- (c5);
\draw[red] (c4) -- (c1);

\end{tikzpicture}
    \caption{A mark path dependent (nonlinear) Hawkes representation of network-update data. Red denotes the current update mark and black the accumulated graph. Because the embedding of an old update can change over time (e.g.\ through triangle participation), its contribution to future intensity need not be fixed at birth.}
    \label{fig:networkHawkes}
\end{figure}

\begin{figure}[!htb]
\centering

\colorlet{mpdcol}{red!55!blue} 
\tikzset{
  highlight/.style={draw=black, line width=0.6pt, minimum width=18pt, minimum height=18pt, inner sep=1pt},
  branch/.style={->, line width=0.9pt},
  histdep/.style={->, mpdcol, line width=0.9pt}
}

\begin{subfigure}[b]{0.47\textwidth}
\centering
\begin{tikzpicture}[>=stealth,scale=0.72,transform shape]
\draw (0,0) -- (10,0);

\node[blue] (leftO) at (0,0) {\Large O};
\node[blue] (x1)    at (2,0) {\Large X};
\node[blue] (x2)    at (3,0) {\Large X};

\node[red]  (midO)  at (5,0) {\Large O};
\node[red]  (rx1)   at (6,0) {\Large X};
\node[red]  (rx2)   at (7,0) {\Large X};

\node[blue,highlight] (boxX) at (9,0) {\Large X};

\draw[branch,blue] (leftO.north) .. controls (1,1.2) .. (x1.north);
\draw[branch,blue] (leftO.north) .. controls (2,1.6) .. (x2.north);
\draw[branch,blue] (leftO.north) .. controls (6,2.2) .. (boxX.north);

\draw[branch,red]  (midO.north)  .. controls (5.5,1.1) .. (rx1.north);
\draw[branch,red]  (midO.north)  .. controls (6.2,1.4) .. (rx2.north);
\end{tikzpicture}
\caption{Linear Hawkes (Poisson cluster view).}
\label{fig:linearHawkes}
\end{subfigure}
\hfill
\begin{subfigure}[b]{0.47\textwidth}
\centering
\begin{tikzpicture}[>=stealth,thick,scale=0.72,transform shape]
\colorlet{purple}{red!55!blue} 
\draw (0,0) -- (10,0);
\node[purple] (leftO) at (0,0) {\Large $\bullet$};
\node[purple] (x1) at (2,0) {\Large $\bullet$};
\node[purple] (x2) at (3,0) {\Large $\bullet$};
\node[purple] (midO) at (5,0) {\Large $\bullet$};
\node[purple] (rx1) at (6,0) {\Large $\bullet$};
\node[purple] (rx2) at (7,0) {\Large $\bullet$};
\node[purple, draw=black, thick, minimum width=18pt, minimum height=18pt] (boxX) at (9,0) {\Large $\bullet$};

\draw[->,purple, opacity=0.8] (leftO.north) .. controls (1,1.2) .. (x1.north);
\draw[->,purple, opacity=0.6] (leftO.north) .. controls (2,1.6) .. (x2.north);
\draw[->,purple, opacity=0.4] (leftO.north) .. controls (3,2.0) .. (midO.north);
\draw[->,purple, opacity=0.3] (leftO.north) .. controls (5,2.4) .. (rx1.north);
\draw[->,purple, opacity=0.2] (leftO.north) .. controls (6,2.6) .. (rx2.north);
\draw[->,purple, opacity=0.1] (leftO.north) .. controls (7,3.0) .. (boxX.north);

\draw[->,purple, opacity=0.9] (x1.north) .. controls (2.5,1.0) .. (x2.north);
\draw[->,purple, opacity=0.6] (x1.north) .. controls (3.5,1.6) .. (midO.north);
\draw[->,purple, opacity=0.5] (x1.north) .. controls (4.5,2.0) .. (rx1.north);
\draw[->,purple, opacity=0.4] (x1.north) .. controls (5.5,2.4) .. (rx2.north);
\draw[->,purple, opacity=0.2] (x1.north) .. controls (6.5,2.8) .. (boxX.north);

\draw[->,purple, opacity=0.8] (x2.north) .. controls (3.5,1.0) .. (midO.north);
\draw[->,purple, opacity=0.6] (x2.north) .. controls (4.5,1.5) .. (rx1.north);
\draw[->,purple, opacity=0.5] (x2.north) .. controls (5.5,1.8) .. (rx2.north);
\draw[->,purple, opacity=0.3] (x2.north) .. controls (6.5,2.2) .. (boxX.north);

\draw[->,purple, opacity=0.9] (midO.north) .. controls (5.5,1.0) .. (rx1.north);
\draw[->,purple, opacity=0.8] (midO.north) .. controls (6.0,1.5) .. (rx2.north);
\draw[->,purple, opacity=0.5] (midO.north) .. controls (7.0,2.0) .. (boxX.north);

\draw[->,purple, opacity=0.9] (rx1.north) .. controls (6.5,1.2) .. (rx2.north);
\draw[->,purple, opacity=0.6] (rx1.north) .. controls (7.5,1.8) .. (boxX.north);

\draw[->,purple, opacity=0.8] (rx2.north) .. controls (8.0,1.2) .. (boxX.north);
\end{tikzpicture}
\caption{Mark path dependent (nonlinear).}
\label{fig:nonlinearHawkes}
\end{subfigure}

\caption{Linear versus mark path dependent Hawkes structure. Left: linear Hawkes admits a cluster genealogy. Right: each event may depend on the full current history, so the arrows indicate history dependence rather than branching ancestry.}

\label{fig:hawkes_linear_vs_nonlinear}
\end{figure}

As discussed in Section~\ref{sec:mpp_notation}, the key ``non-separability'' phenomenon in network growth is topology timing feedback: the current graph state can influence how soon the next update occurs, and the temporal regime of activity can influence which updates are likely. This feedback is especially salient when excitation depends on motifs (triangles, $k$-stars, shared partners), because the embedding of an old edge can change as new edges arrive. We therefore allow the excitation kernel to depend on the current marked history.

\begin{definition}[Mark path dependent Hawkes process]\label{def:mpd_hawkes}
Let $N$ be a simple marked point process on $\mathbb{R}_+ \times \mathcal{M}$ with natural filtration $\{\mathcal{H}_t\}_{t \ge 0}$. We say that $N$ is a mark path dependent Hawkes process if its $\mathcal{H}_t$-conditional intensity admits the representation
\begin{equation}
\lambda(t,m \mid \mathcal{H}_t)
=
\lambda_{\emptyset}(t,m)
+
\int_{(0,t)\times\mathcal{M}}
g\!\left(t-u, m, m' \mid \mathcal{H}_t\right)\,
N(du,dm'),
\label{eq:path_dependent_marked_hawkes}
\end{equation}
where $\lambda_{\emptyset}(t,m)$ is a predictable baseline intensity and $g$ is a predictable excitation kernel.
\end{definition}

The key difference from Definition~\ref{def:mpp_hawkes} is that the contribution of a past update need not be fixed at its birth: it may vary with the evolving graph state. More concretely, the excitation kernel depends on the full history (including mark path). This non linearity is not an external ``activation function'' applied uniformly to the integrated past as discussed in \cite{bremaud1996stability}. Despite the fact that the contribution of a past update need not be fixed at its birth (and evolves with respect to graph state), a process characterised by Definition~\ref{def:mpp_hawkes} does not violate causality. At time $t$ the intensity is evaluated using the information available at time $t-$, and later network growth may change the topological role of earlier nodes and edges.

For likelihood-based inference, the model must define a well-posed stochastic process. 
For mark path dependent Hawkes processes, cluster-based arguments are not available in general, so we work with the Poisson-embedding (thinning) representation and a contraction argument in a natural metric on marked histories. Intuitively, the key requirement is that changing the past by adding or removing a single event can only affect the current intensity by an amount that decays with that event’s age, and that these effects are sufficiently summable over time. This is precisely what explicit time decay and bounded local update mechanisms enforce in the induced network models in Section~\ref{sec:induced_models}.

\subsection{Poisson embedding and simulation via thinning}\label{sec:poisson_embedding}

The Poisson embedding is the common engine behind (i) the fixed-point definition of the model, (ii) existence/uniqueness and non-explosion proofs via contraction, and (iii) exact simulation by thinning. The logic is as follows: specify an intensity functional $\eta\mapsto \lambda(\cdot,\cdot\mid\eta)$ on marked histories, embed it into a single Poisson random measure, and then construct the process as a self-consistent thinning of that Poisson measure \citep{bremaud1996stability}.

Let $\Pi$ be a Poisson random measure on $E=\mathbb R\times\mathbb R_+\times\mathcal M$ with intensity $dt\,dz\,\nu(dm)$, where $\nu$ is counting measure on $\mathcal M$. Given a predictable marked intensity $\lambda(t,m\mid\mathcal H_t)$, define a marked point process by thinning:
\begin{equation}\label{eq:thinning}
N(dt,dm)=\int_{\mathbb R_+}\mathbf 1\{z\le \lambda(t,m\mid\mathcal H_t)\}\,\Pi(dt,dz,dm).
\end{equation}
Equation \eqref{eq:thinning} is both the stochastic fixed-point equation defining the process and (after choosing a dominating proposal rate) the basis for simulation by thinning.

We state assumptions directly on the full marked intensity functional $\eta\mapsto \lambda(\cdot,\cdot\mid\eta)$, where $\eta$ is a generic marked history (an integer-valued measure on $\mathbb R\times\mathcal M$). Throughout Section~\ref{sec:poisson_embedding}, a ``history'' $\eta$ denotes a locally finite integer-valued measure on
$\mathbb R_+\times\mathcal M$, i.e.
\[
\eta([0,T]\times\mathcal M)<\infty \qquad \text{for all }T<\infty.
\]
We write $\eta_t$ for the restriction of $\eta$ to $[0,t)\times\mathcal M$. 

To state a contraction condition in the marked setting, we measure discrepancies between histories in a weighted total-variation norm, so that larger updates count as larger perturbations. We now proceed to define a weighted metric on marked histories. Fix a measurable weight function $w:\mathcal M\to[1,\infty)$ (e.g.\ $w(m)=1+|m\cap\mathcal V|+|m\cap\mathcal E|$). For any signed measure $\mu$ on $\mathbb R_+\times\mathcal M$, define the $w$-weighted total variation measure by
\[
|\mu|_w(ds,dm):=w(m)\,|\mu|(ds,dm).
\]
For two integer-valued measures $\eta,\eta'$ we write
\[
|\eta-\eta'|_w(ds,dm):=w(m)\,|\eta-\eta'|(ds,dm),
\qquad
\|\eta-\eta'\|_{w,T}:=|\eta-\eta'|_w([0,T)\times\mathcal M).
\]
When $\eta,\eta'$ are integer-valued, $\|\eta-\eta'\|_{w,T}$ equals the sum of weights $w(m)$ over atoms that belong to exactly one of $\eta$ or $\eta'$ on $[0,T)\times\mathcal M$.

\begin{assumption}[Bounded baseline]
\label{ass:H0}
Let $\mathbf 0$ denote the empty history. Define
\[
\lambda_{\emptyset}(t,m):=\lambda(t,m\mid\mathbf 0),
\qquad
\lambda_{g,\emptyset}(t):=\sum_{m\in\mathcal M} \lambda_{\emptyset}(t,m),
\qquad
\lambda_{g,\emptyset}^{(w)}(t):=\sum_{m\in\mathcal M} w(m)\,\lambda_{\emptyset}(t,m).
\]
Assume
\[
\bar\lambda_\emptyset^{(w)} := \sup_{t\ge 0}\lambda_{g,\emptyset}^{(w)}(t) <\infty.
\]
(Then $\sup_{t\ge0}\lambda_{g,\emptyset}(t)\le \bar\lambda_\emptyset^{(w)}$ since $w\ge1$.)
\end{assumption}

\begin{assumption}[Weighted marked Lipschitz contraction]\label{ass:HL}
There exists measurable $h:\mathbb R_+\to\mathbb R_+$ with
\[
\|h\|_1:=\int_0^\infty h(u)\,du<1
\]
such that for all $t\ge 0$ and all histories $\eta,\eta'$ (locally finite integer-valued measures on $\mathbb R_+\times\mathcal M$),
\begin{equation}\label{eq:HL_lambda_marked}
\sum_{m\in\mathcal M} w(m)\,
\bigl|\lambda(t,m\mid\eta_t)-\lambda(t,m\mid\eta'_t)\bigr|
\le
\int_{[0,t)\times\mathcal M} h(t-s)\,|\eta-\eta'|_w(ds,dm),
\end{equation}
where $|\eta-\eta'|_w(ds,dm):=w(m)\,|\eta-\eta'|(ds,dm)$ is the weighted total variation measure.
\end{assumption}

\begin{remark}[Evaluating $\lambda$ on counterfactual histories]
\label{rem:counterfactual_domain}
In Assumptions~\ref{ass:H0}--\ref{ass:HL} (and in the one-event influence definition below), the argument $\eta$ denotes an arbitrary locally finite integer-valued measure on $\mathbb R_+\times\mathcal M$, not necessarily a realizable network history under
Definition~\ref{def:admissible_marks}. To ensure $\lambda(t,m\mid \eta_t)$ is well-defined on this larger domain (including for the counterfactual histories obtained by adding/removing atoms in telescoping arguments) we interpret any $\eta_t$ as generating a graph state via the deterministic accumulation rule
\[
V(\eta_t)
:= V_0 \cup \bigcup_{(s,m)\in \eta_t}\Bigl((m\cap\mathcal V)\cup \operatorname{end}(m\cap\mathcal E)\Bigr),
\qquad
E(\eta_t)
:= E_0 \cup \bigcup_{(s,m)\in \eta_t}(m\cap\mathcal E),
\]
where $\operatorname{end}(\Delta E):=\bigcup_{e\in\Delta E} e$ denotes the set of endpoints of edges in $\Delta E$.
We then define the admissible mark set at time $t$ under history $\eta$ as
\[
\mathcal M_t(\eta):=\{m\in\mathcal M:\ m \text{ is admissible relative to } (V(\eta_t),E(\eta_t))
\text{ in the sense of Definition~\ref{def:admissible_marks}}\},
\]
and enforce admissibility by $\lambda(t,m\mid\eta_t)=0$ for $m\notin \mathcal M_t(\eta)$.
For histories $\eta=N$ generated by the HawkesNet process, this extension coincides with the intended
network state $\mathcal G_{t-}$ and admissible set $\mathcal M_t$.
\end{remark}

\begin{theorem}[Existence, pathwise uniqueness, non-explosion]\label{thm:existence}
Assume \ref{ass:H0}--\ref{ass:HL}. Then there exists a pathwise unique simple marked point process $N$ on $[0,\infty)\times\mathcal M$ satisfying the fixed-point thinning equation \eqref{eq:thinning}. Moreover, for every $T>0$, $\mathbb E[N([0,T]\times\mathcal M)]<\infty$, hence $N([0,T]\times\mathcal M)<\infty$ almost surely.
\end{theorem}

The proof couples Picard iterates through a common Poisson random measure and applies a thinning-coupling inequality together with a contraction argument on marked histories; full details are deferred to Appendix~\ref{sec:stability_theory}. For induced HawkesNet models, the same appendix also gives a simpler predictable-marking route: if a non-explosive ground process is combined with a predictable network-update PMF, then the resulting marked process is well posed.

\begin{assumption}[Bounded update size]
\label{ass:bounded_update}
There exist constants $B_V,B_E<\infty$ such that for all $t$ and all admissible marks $m\in\mathcal M_t$,
\[
|m\cap\mathcal V|\le B_V,
\qquad
|m\cap\mathcal E|\le B_E.
\]
That is, each update adds at most $B_V$ new nodes and at most $B_E$ new edges.
\end{assumption}

\begin{corollary}[Bounded mean activity and at-most-linear expected network growth]
\label{cor:linear_growth}
Assume \ref{ass:H0}--\ref{ass:HL} and \ref{ass:bounded_update}.
Let $N_g([0,T]):=N([0,T]\times\mathcal M)$ denote the number of updates on $[0,T]$ (i.e.\ the ground count).
Then there exists a constant $C_\lambda<\infty$ such that for all $T>0$,
\[
\mathbb E\bigl[N_g([0,T])\bigr]\le C_\lambda\,T.
\]
Consequently, the expected node and edge counts of the cumulative graph satisfy
\[
\mathbb E[N_T^{\text{nodes}}]\le N_0 + B_V\,C_\lambda\,T,
\qquad
\mathbb E[|E_T|]\le |E_0| + B_E\,C_\lambda\,T,
\]
i.e.\ cumulative network size grows at most linearly in expectation.

Moreover, let \(a:\mathbb R_+\to\mathbb R_+\) be an aging kernel with \(\|a\|_1<\infty\) and \(\|a\|_\infty<\infty\). Define the aged edge mass
\[
|E_t|^{(a)}:=\sum_{(i,j)\in E_t} a(t-b_{i,j}).
\]
Then
\[
\sup_{t\ge 0}\mathbb E[|E_t|^{(a)}] < \infty,
\]
so time-local (aged) network totals remain uniformly controlled under the same assumptions.
\end{corollary}

\subsubsection{Simulation via thinning}\label{sec:simulation}

The same construction yields exact simulation by thinning \citep{ogata1981lewis}. Writing
\[
\lambda_g(t\mid\mathcal H_t):=\sum_{m\in\mathcal M}\lambda(t,m\mid\mathcal H_t),
\]
we choose a dominating rate $\Lambda(t)\ge \lambda_g(t\mid\mathcal H_t)$, propose candidate times from $\Lambda$, and accept a candidate time $t^*$ with probability
\[
\frac{\lambda_g(t^*\mid\mathcal H_{t^*})}{\Lambda(t^*)}.
\]
Conditional on acceptance, the mark is drawn from
\[
\mathbb P(M=m\mid t^*,\mathcal H_{t^*})
=
\frac{\lambda(t^*,m\mid\mathcal H_{t^*})}{\lambda_g(t^*\mid\mathcal H_{t^*})}.
\]
In the induced class $\lambda=\lambda_g q$, this final step reduces to sampling from the network-update PMF $q$. Full pseudocode is given in Algorithm~\ref{alg:ogata_thinning} of Appendix~\ref{sec:sim_algorithm}.


\subsubsection{Design principles for stability in induced HawkesNet models}

Stability assumptions in Section~\ref{sec:poisson_embedding} are imposed on the full marked intensity $\lambda(t,m\mid\mathcal H_t)$, but in induced HawkesNet models it is often easiest to reason through two design principles that appear repeatedly in network growth mechanisms: decaying in time, and saturation. 

If the effect of a past event on the current intensity decays with its age, then adding or removing a single past update can only change the present by an amount that shrinks as that update recedes into the past. This is exactly what makes a kernel $h(\cdot)$ integrable and small in $L^1$, the key requirement for contraction. In induced models, decay can appear in the ground rate (e.g.\ Hawkes kernels) and/or in the mark PMF (e.g.\ $\exp(-\tau(t-t_i))$ age-weighting in attachment probabilities).

To obtain time-local summaries that remain interpretable and to support stability arguments, we define aged(finite-memory) versions of network features. Many network features used for update probabilities (motif counts, local degrees, triangle participation) can grow with network size. If the mark PMF responds linearly to such features, small changes in history can lead to large changes in intensity, potentially violating the contraction condition and enabling supercritical cascades. Using saturating links (e.g.\ logistic transforms), truncating/extending statistics in a bounded way, or restricting updates to local neighborhoods keeps the incremental effect of any single past update uniformly controlled.

These principles are not just theoretical conveniences, they ensure that the process is well-posed and therefore likelihood based inference is feasible. In practice one can (i) enforce decay through parametric restrictions (e.g.\ $\beta>0$ and moderate $K/\beta$ in the ground rate) and (ii) enforce bounded sensitivity by using probability models for marks that are inherently bounded (e.g.\ Bernoulli/logistic links) and by avoiding unbounded linear amplification of motif counts. Lemma~\ref{lem:HL_no_uniform_envelope} below gives a convenient route to verify Assumption~\ref{ass:HL} under the factorization $\lambda=\lambda_g q$ without assuming a uniform envelope on $\lambda_g$.

\begin{lemma}[Route to Assumption~\ref{ass:HL} under $\lambda=\lambda_g q$]
\label{lem:HL_no_uniform_envelope}
Assume $\lambda(t,m\mid\eta_t)=\lambda_g(t\mid\eta_t)\,q(m\mid t,\eta_t)$ where $q(\cdot\mid t,\eta_t)$ is a PMF on $\mathcal M$.
Define the conditional mean weight
\[
\mu_w(t,\eta_t):=\sum_{m\in\mathcal M} w(m)\,q(m\mid t,\eta_t),
\qquad
\bar\mu_w:=\sup_{t\ge0}\sup_{\eta}\mu_w(t,\eta_t).
\]
Assume $\bar\mu_w<\infty$. Suppose there exist measurable kernels $h_g,h_{q,\lambda}:\mathbb R_+\to\mathbb R_+$ such that for all $t$ and
histories $\eta,\eta'$,
\begin{align*}
|\lambda_g(t\mid\eta_t)-\lambda_g(t\mid\eta'_t)|
&\le \int_{[0,t)\times\mathcal M} h_g(t-s)\,|\eta-\eta'|_w(ds,dm),\\
\sum_{m\in\mathcal M}\lambda_g(t\mid\eta'_t)\,w(m)\,\bigl|q(m\mid t,\eta_t)-q(m\mid t,\eta'_t)\bigr|
&\le \int_{[0,t)\times\mathcal M} h_{q,\lambda}(t-s)\,|\eta-\eta'|_w(ds,dm).
\end{align*}
Then Assumption~\ref{ass:HL} holds with $h:=\bar\mu_w\,h_g+h_{q,\lambda}$.
\end{lemma}

Assumption~\ref{ass:HL} is a sufficient condition for existence/uniqueness/non-explosion of fully coupled marked intensities
$\eta\mapsto \lambda(\cdot,\cdot\mid\eta)$, where the marks can feed back into the future ground rate.
In contrast, for the induced HawkesNet models in Section~\ref{sec:induced_models} we often specify a well-posed ground process $N_g$
(e.g.\ a classical unmarked Hawkes process) and then attach marks via a predictable network-update PMF $q(\cdot\mid t,\mathcal H_t)$.
In this setting the marked process is well-posed whenever the ground process is non-explosive, without requiring the uniform-in-history contraction \eqref{eq:HL_lambda_marked}; see Proposition~\ref{prop:induced_marking_wellposed} in Appendix~\ref{sec:induced_marking_wellposed}. This applies directly to the BA and CS mechanisms in Examples~\ref{ex:BA}--\ref{ex:cs}.

\subsubsection{One-event influence and a sharp contraction envelope}
\label{sec:one_event_influence}

Assumption~\ref{ass:HL} posits the existence of an envelope $h$ controlling how changes in the past
perturb the present marked intensity in $\ell^1(\mathcal M)$.
A convenient way to diagnose when such an envelope can exist is to quantify the worst-case effect of a single update.

\begin{definition}[Maximal one-event influence kernel
]
\label{def:one_event_influence}
For $u>0$, define the weighted one-event influence at lag $u$ by
\begin{equation}
\label{eq:alpha_def}
\alpha_w(u)
:=
\sup_{t\ge u}\;
\sup_{\eta}\;
\sup_{m_0\in\mathcal M}\;
\frac{1}{w(m_0)}
\sum_{m\in\mathcal M}
w(m)\Bigl|
\lambda\!\left(t,m \mid \eta_t+\delta_{(t-u,m_0)}\right)
-
\lambda\!\left(t,m \mid \eta_t\right)
\Bigr|,
\end{equation}
where the supremum over $\eta$ ranges over locally finite integer-valued measures on
$\mathbb R_+\times\mathcal M$, and $\eta_t$ denotes restriction to $[0,t)\times\mathcal M$.
\end{definition}

Whenever we invoke an integrability condition such as $\int_0^\infty \alpha_w(u)\,du<\infty$, we additionally assume that $u\mapsto \alpha_w(u)$ is measurable. This is automatic under mild regularity conditions on $\eta\mapsto \lambda(t,\cdot\mid \eta_t)$,
but we state it explicitly since $\alpha$ is defined via a supremum.

\begin{lemma}[One-event influence yields a sharp HL envelope]
\label{lem:alpha_implies_HL}
Assume $\alpha_w(u)<\infty$ for all $u>0$. Then for all $t\ge 0$ and histories $\eta,\eta'$,
\begin{equation}
\label{eq:HL_from_alpha}
\sum_{m\in\mathcal M}
w(m)\bigl|\lambda(t,m\mid\eta_t)-\lambda(t,m\mid\eta'_t)\bigr|
\le
\int_{[0,t)\times\mathcal M}\alpha_w(t-s)\,|\eta-\eta'|_w(ds,dm).
\end{equation}
Consequently, Assumption~\ref{ass:HL} holds with $h=\alpha_w$ whenever $\alpha_w$ is measurable and
$\int_0^\infty \alpha_w(u)\,du<1$.

Moreover, if Assumption~\ref{ass:HL} holds for some measurable kernel $h$, then
\begin{equation}
\label{eq:alpha_le_h}
\alpha_w(u)\le h(u)\qquad\text{for all }u>0,
\end{equation}
so $\alpha_w$ is pointwise minimal among all envelopes that can satisfy \eqref{eq:HL_lambda_marked}.
\end{lemma}

\begin{corollary}[Well-posedness via one-event influence (weighted)]
\label{cor:alpha_wellposed}
If Assumption~\ref{ass:H0} holds, $\alpha_w$ is measurable, and $\int_0^\infty \alpha_w(u)\,du<1$, then
Assumptions~\ref{ass:H0}--\ref{ass:HL} hold (with $h=\alpha_w$), and therefore Theorem~\ref{thm:existence} applies.
\end{corollary}

If $\alpha_w(u)=\infty$ for some $u>0$, then no integrable $h$ can satisfy Assumption~\ref{ass:HL} (since any such $h$ must dominate $\alpha_w$ by \eqref{eq:alpha_le_h}). This identifies a structural obstruction: if a single early update can later have unbounded worst-case effect on network statistics driving the intensity (e.g.\ through globally growing motif participation), then a uniform-in-history contraction criterion cannot hold without additional design constraints such as aging, localization, or saturation.


\subsection{Likelihood and Parameter Estimation}\label{sec:likelihood}

Throughout, we estimate parameters by maximizing the standard marked point process log-likelihood \citep{daley2003introduction,rathbun1996asymptotic}. Under the non-explosion guarantee of Theorem~\ref{thm:existence}, the compensator is finite on finite observation windows, and the likelihood is therefore well-defined. Our focus in this paper is on (i) specifying a flexible non-separable model class for growing networks, (ii) enabling tractable likelihood evaluation via the PMF construction, and (iii) demonstrating reliable finite-sample behavior through simulation and applications. As is common in likelihood-based point process modeling \citep{schoenberg2019recursive}, the contributions of this paper do not depend on a complete large-$T$ asymptotic theory (consistency/CLT for estimators under appropriate regularity conditions). Establishing such results, along the lines of  \citet{ogata1978estimators}, is an important complementary direction for future theoretical work.

Let $\theta$ denote the full parameter vector.
Given observed updates $\{(t_i,m_i)\}_{i=1}^n$ on $[0,T)$, the log-likelihood for a simple marked point process with
$\mathcal H_t$-conditional intensity $\lambda_\theta(t,m\mid\mathcal H_t)$ (with respect to $dt\,\nu(dm)$) is
\begin{equation}
\label{eq:loglik_general}
\ell(\theta)
=
\sum_{i=1}^n \log \lambda_\theta(t_i,m_i\mid\mathcal H_{t_i})
-
\int_0^T \left(\int_{\mathcal M} \lambda_\theta(t,m\mid\mathcal H_t)\,\nu(dm)\right)\,dt .
\end{equation}
$\mathcal M$ is countable and $\nu$ is counting measure, so the compensator term is
$\int_0^T \sum_{m\in\mathcal M} \lambda_\theta(t,m\mid\mathcal H_t)\,dt$.
Since $\lambda_\theta(t,m\mid\mathcal H_t)=0$ for $m\notin\mathcal M_t$, the sum may be taken over $\mathcal M_t$.

For a mark path dependent Hawkes process (Definition~\ref{def:mpd_hawkes}) the kernel $g_\theta$ may depend on both the candidate mark $m$ and the past mark $m_j$ (in addition to time and the current history). In this case, the compensator
\[
\int_0^T \sum_{m\in\mathcal M_t}
\left\{\lambda_{\emptyset,\theta}(t,m)+\sum_{j:\,t_j<t} g_\theta(t-t_j,m,m_j\mid\mathcal H_t)\right\}dt
\]
is typically intractable because $|\mathcal M_t|$ is enormous and history-dependent. This is remedied with the algebraic decomposition in Equation \ref{eq:algebraic_factorization}

Admissibility is enforced by $\lambda_\theta(t,m\mid\mathcal H_t)=0$ for $m\notin\mathcal M_t$. Equation \eqref{eq:algebraic_factorization} is not an assumption and does not impose mark separability: both $\lambda_{g,\theta}$ and $q_\vartheta$ may depend on the full marked history through the evolving graph (see Remark~\ref{rem:topology_timing_vs_decomp}). What \eqref{eq:algebraic_factorization} provides is a likelihood identity: since $q_\theta(\cdot\mid t,\mathcal H_t)$ is a PMF, $\sum_{m\in\mathcal M} q_\theta(m\mid t,\mathcal H_t)=1$, and therefore the compensator reduces to
\begin{equation}
\label{eq:compensator_reduction}
\int_0^T \sum_{m\in\mathcal M} \lambda_\theta(t,m\mid\mathcal H_t)\,dt
=
\int_0^T \lambda_{g,\theta}(t\mid\mathcal H_t)\,dt.
\end{equation}
Substituting into \eqref{eq:loglik_general} yields the decomposition
\begin{equation}
\label{eq:loglik_factorized}
\ell(\theta)
=
\sum_{i=1}^n \log q_\theta(m_i\mid t_i,\mathcal H_{t_i})
+
\sum_{i=1}^n \log \lambda_{g,\theta}(t_i\mid \mathcal H_{t_i})
-
\int_0^T \lambda_{g,\theta}(t\mid\mathcal H_t)\,dt.
\end{equation}

While \eqref{eq:algebraic_factorization} and \eqref{eq:loglik_factorized} hold for all specifications,
they are not automatically computationally helpful: in the fully general model \eqref{eq:path_dependent_marked_hawkes},
evaluating $\lambda_{g,\theta}(t\mid\mathcal H_t)=\sum_{m\in\mathcal M_t}\lambda_\theta(t,m\mid\mathcal H_t)$ may still require an infeasible sum over the dynamic mark space.
To obtain tractable likelihood evaluation we therefore model directly in the form \eqref{eq:algebraic_factorization} by specifying $q_\theta(\cdot\mid t,\mathcal H_t)$ as a normalized network-update PMF and $\lambda_{g,\theta}$ as a separate, explicit ground-rate model. In this induced HawkesNet class, the candidate-mark dependence of $\lambda_\theta(t,m\mid\mathcal H_t)$ enters only through $q_\theta(\cdot\mid t,\mathcal H_t)$, while the ground rate $\lambda_{g,\theta}(t\mid\mathcal H_t)$ may still depend on the marked history (and hence the current network state). This preserves topology--timing feedback without requiring an explicit summation over $\mathcal M_t$ in the compensator.

We model the ground rate with a Hawkes-type kernel,
\begin{equation}
\label{eq:ground_hawkes_general}
\lambda_{g,\theta}(t\mid\mathcal H_t)
=
\lambda_{\emptyset}
+
\sum_{j:\,t_j<t} \phi_\theta(t-t_j,m_j\mid\mathcal H_t),
\end{equation}
where $\phi_\theta$ may depend on past marks and the evolving graph state (through $\mathcal H_t$),
but does not depend on the candidate mark $m$.
For maximum computational efficiency we often take a purely temporal exponential kernel
$\phi_\theta(t-t_j,m_j\mid\mathcal H_t)=K\exp(-\beta(t-t_j))$, giving
\[
\lambda_{g}(t)=\lambda_{\emptyset} + K\sum_{j:\,t_j<t} e^{-\beta(t-t_j)}.
\]
In this exponential (Markovian) case the compensator admits the closed form
\begin{equation}
\label{eqn:simpleLL}
\int_0^T \lambda_g(t)\,dt
=
\lambda_{\emptyset}\,T
+
\frac{K}{\beta}\sum_{j=1}^{n}\left(1-e^{-\beta(T-t_j)}\right),
\end{equation}
which yields an $\mathcal O(n)$ likelihood evaluation when computed in its recursive form, further facilitating practical maximum likelihood estimation \citep{cui2020elementary}. Other common choices include power law, Gamma, and Mittag-Leffler kernels \citep{hawkes2018hawkes}.

Although the likelihood in \eqref{eq:loglik_factorized} separates into a mark term and a ground-rate term,
the model remains non-separable in the substantive sense relevant for evolving networks: both components may depend on the marked history through the current network state. Temporal parameters should therefore be interpreted as governing the update-time intensity conditional on the evolving network, rather than conditional on a proposed candidate mark.
In practice we maximize \eqref{eq:loglik_factorized} numerically; in our examples we use Nelder--Mead \citep{NelderMead1965}
via \texttt{optim} in R \citep{RCoreTeam2025} as implemented in our \texttt{HawkesNet} package 
\if0\anon (citation blinded for peer review) \else \citep{hawkesNet}
\fi. 

Finally, we propose two options for recovering standard errors of parameter estimates: (i) parametric bootstrap allows for reporting empirically variability of bootstrap estimates (via thinning as described in Section~\ref{sec:simulation}; this approach underlies our simulation-based goodness-of-fit diagnostics in Section~\ref{sec:hypertext_gof}), or (ii) assuming asymptotic normality and using the Hessian (as is standard in the point process literature, see for instance \cite{schoenberg2019recursive}). We note that even in settings where the asymptotic theory is well developed for univariate Hawkes \cite{ogata1978estimators} and univariate marked Hawkes processes \cite{rathbun1996asymptotic}, likelihood surfaces are often flat or ridgy, and therefore asymptotic standard errors can be unreliable \cite{schoenberg2016note,kresin2023parametric}.

\section{Simulation Study}\label{sec:simulation_study}
We evaluate finite-sample behavior of maximum likelihood estimation in the induced HawkesNet class under two update mechanisms: preferential-attachment marks (BA; Example~\ref{ex:BA}) and change-statistic marks (CS; Example~\ref{ex:cs}). For each design we simulate $100$ independent realizations on a fixed window $[0,T)$ using Ogata-style thinning (Section~\ref{sec:simulation}; Algorithm~\ref{alg:ogata_thinning} in Appendix~\ref{sec:sim_algorithm}), then refit the model by maximizing the marked point process likelihood as in Section~\ref{sec:likelihood}. Full design choices (including parameter settings, candidate-set restrictions for CS, and any fixed-parameter identifiability choices) are reported in Appendix~\ref{app:sim_study}, with numerical summaries in Tables~\ref{tab:BA_table} and~\ref{tab:CS_table}.

Across both mechanisms, mark PMF parameters are recovered accurately in stable regimes, while Hawkes temporal parameters exhibit the familiar finite-sample variability of self-exciting models. Additional diagnostics, including sampling distributions of the MLEs (Appendix~\ref{sec:mle_dist}), increasing-$T$ consistency checks (Appendix~\ref{sec:app_consistency}), and an illustration of the transition to high growth behavior in critical and supercritical settings are provided in the appendix (Appendix~\ref{sec:app_explosivity_BA}).

\section{Application to Dynamic Human Contact Patterns}\label{sec:hypertext}

We apply HawkesNet to a dynamic face-to-face contact network recorded during the first day of the ACM Hypertext 2009 conference \citep{hypertext_conf_1, sociopatterns2025}. Conference interactions naturally exhibit self-exciting, non-separable dynamics: an initial contact often sparks subsequent localized interactions as participants join ongoing conversations or are drawn into shared social contexts. We model the monotonic growth of this network over its primary 16-hour session, comprising $n=100$ unique attendees and 946 first-contact edges. Full details regarding data preprocessing, continuous-time jittering, and network construction are provided in Appendix \ref{sec:supp_data_processing}. The raw data are publicly available at \url{https://www.sociopatterns.org/datasets/hypertext-2009-dynamic-contact-network/}.

\subsection{Model specification}\label{sec:hypertext_model}

We fit the change-statistic (CS) HawkesNet model as described in Example~\ref{ex:cs}. We found star and triangle count models did not recreate the observed node degree distribution; see Appendix-\ref{app:bad_fit} for details. We follow the ERGM literature \citep{snijders2006} and include the geometrically weighted edgewise shared partner (GWESP) terms and the geometrically weighted degree (GWDEG) terms, defined as follows:
\begin{align*}
\text{GWESP}(\alpha) = \sum_{i=1}^{n-2} e^{-\alpha (i-1)} ESP_{i} \quad\text{and}\quad
\text{GWDegree}(\alpha) = \sum_{k=1}^{n-1} e^{-\alpha (k-1)} DEG_{k}
\end{align*}
Where $ESP_{i}$ is the number of edges with $i$ shared partners and $DEG_{k}$ is the number of nodes with degree $k$. The GWESP and GWDEG terms, model a similar process to the additional transitivity and super popularity of some nodes, above that expected from independent edges. However, they allow for higher order transitivity, and limit explosion to full networks \citep{handcock2003}. In particular, the preference for social closure, yet the lack of explosion to a full network is simply too hard for a single triangle and two star terms to model effectively, hence the poor fit and misleading results shown in Appendix \ref{app:bad_fit}.

The $\lambda_{nodes}$ node addition term and $K$ terms are held fixed at the observe nodes-per-event rate and $1$ respectively to stabilize estimation, consistent with the simulation studies in Section~\ref{sec:CS_sim}.
\subsection{Results}\label{sec:hypertext_results} Table~\ref{tab:ht_fit1} reports the parameter estimates. First, we consider the parameter $K$. While fixing $K = 1$ would be concerning in a standard non-marked path dependent Hawkes process, in our case $K$ cannot be interpreted simply as the expected number of points triggered by a parent. This is due to its non-separable interaction with both the mark PMF and the time-evolving mark space. However, we note that the $\lambda_{\emptyset}$ estimate suggests an expected number of background events of 844 in the $[0,1]$ time interval. Given that we observed 946 points, and simulations from the fitted model produce a similar count, it is clear that under the fitted parameters the model does not reduce to a history-independent process; triggering is actively occurring.

The $\beta_{\text{overall}}$ parameter suggests a rapid decay of event triggering. Given our 16-hour time window, if this were a separable Hawkes model, the expected time of a triggered point would be $\frac{1}{11.734} \approx 1.4$ hours. Furthermore, $\beta_{\text{edges}}$ is of a similar magnitude, suggesting that new edges are preferentially formed between nodes that were last active at similar times. Consequently, inactive nodes become increasingly less ``attractive'' for new edges the longer they remain dormant, which is intuitively reasonable for a conference setting.

The parameter $m$ is estimated as $1$, reflecting the fact that each event added exactly one edge during our data processing step. The edge parameter accounts for the low baseline propensity for edge formation. Notably, both the GWDEG and GWESP terms are positive and statistically significant. This suggests that both social closure and popularity processes are at play. This aligns with the expected social mechanisms of a conference, where edges form as individuals are drawn into small groups of mutual connections (triadic closure) or seek out interactions with highly influential or sociable nodes (preferential attachment).

\subsection{Goodness of Fit}\label{sec:hypertext_gof}

To assess joint network and time GOF we pursue a simulation based approach by simulating $100$ networks from the fitted model and comparing observed and simulated network statistics \citep{Hunter_2008}. GOF diagnostics focus on three structural aspects of the network: the degree distribution, the edgewise shared partner (ESP) distribution and geodesic distances (i.e. shortest paths between node pairs). We show the waiting time distributions for the ESP and degrees distributions. The waiting times combine network structure and time and are therefore well suited to our setting. See \ref{app:RCT_GOF} for additional GOF comments.

This is a high standard for goodness-of-fit for network structure. The ability to recreate the full degree, ESP and geodesic distance distribution with a handful of parameters, suggests the model is an excellent, and parsimonious fit for the data. 

\begin{figure}[htbp]
  \centering

  \begin{subfigure}[t]{0.49\textwidth}
    \centering
    \includegraphics[width=\linewidth]{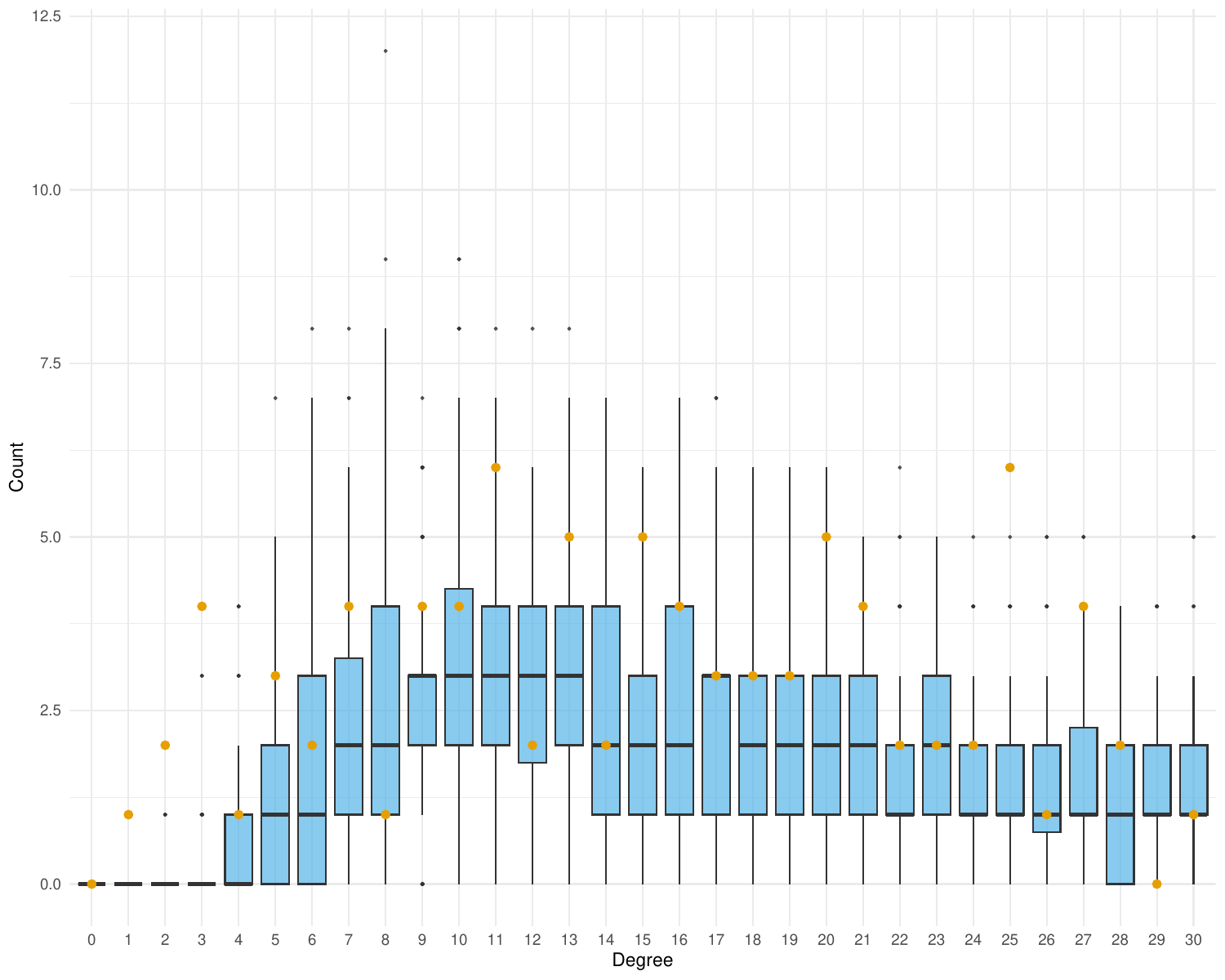}
    \caption{Degree distribution.}
    \label{fig:ht_gof_degree}
  \end{subfigure}\hfill
  \begin{subfigure}[t]{0.49\textwidth}
    \centering
    \includegraphics[width=\linewidth]{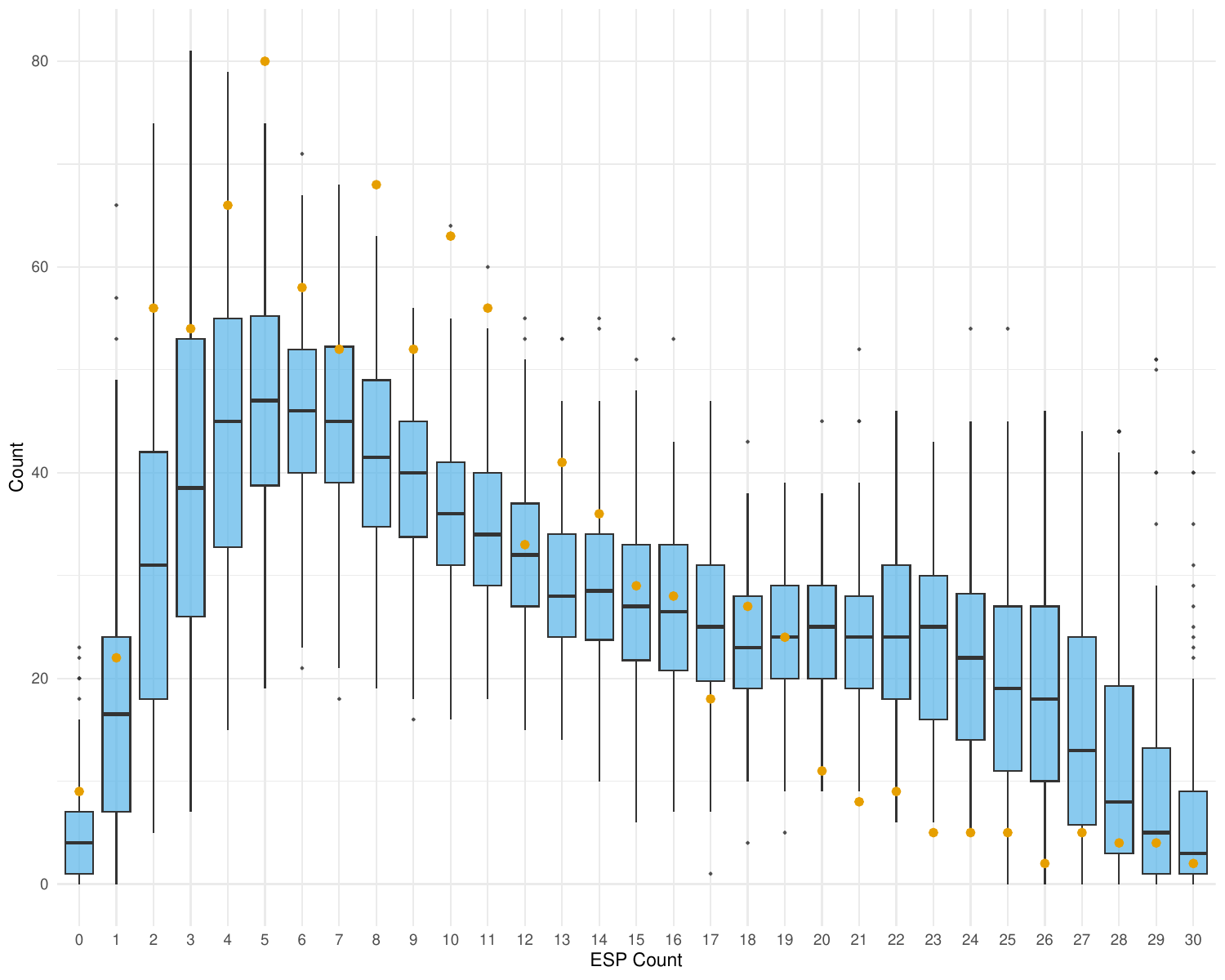}
    \caption{Edgewise shared partner distribution.}
    \label{fig:ht_gof_esp}
  \end{subfigure}

  \caption{GOF diagnostics comparing observed (orange points) vs.\ simulated (blue boxplots): (a) degree distribution and (b) edgewise shared partner distribution.}
  \label{fig:ht_gof}
\end{figure}

Figure \ref{fig:ht_gof_degree} shows the distribution across simulation, of node degree counts, with the observed data plausible from the simulations. The fitted model produces fewer very low degree (1,2,3) nodes than observed, but otherwise fits well. Figure \ref{fig:ht_gof_esp} shows a reasonable fit on ESP with our fitted model slightly under representing lower ESP values and over representing higher ESP values. Figure \ref{fig:ht_gof_geodist} shows similarly plausible geodesic distances between nodes.

Figures \ref{fig:ht_gof_waiting_degs} and \ref{fig:ht_gof_waiting_esp} show the waiting time distributions for both the observed data and the simulations. Again, we note the clear similarity between the observed and simulated data. 

\section{Discussion}\label{sec:discuss}

We introduced HawkesNet, a marked point process framework for continuous-time network evolution. By encoding discrete network updates as marks, our formulation yields a single conditional intensity characterizing both \textit{when} and \textit{how} a network changes. This allows for non-separable dependence between update timing and network topology: even when the ground component is Hawkesian, the effective event rate depends on the network history through the dynamic, evolving mark space. As demonstrated through our preferential attachment and change-statistic specifications, this modular framework provides an interpretable, likelihood-based alternative to descriptive models, successfully capturing both temporal bursting and localized structural mechanisms. 

Our formulation broadens the scope of nonlinear Hawkes processes by introducing mark-path dependence through the evolving graph state: even when the ground component is Hawkesian, the effective event rate depends on the network history through the mark model. This feature is essential for representing growth processes in which the plausibility of updates depends on current topology and also changes how temporal parameters must be interpreted.

While our stability results establish well-posedness for marked, history-dependent settings (Appendix~\ref{sec:stability_theory}), developing sharper, model-specific stability conditions and large-window asymptotic theory remains an important direction for future work. Methodologically, the current implementation focuses on monotonic network growth. Natural extensions involve expanding the mark space to accommodate edge dissolution (birth-death marks), repeated dyadic interactions, and weighted updates. Furthermore, integrating exogenous nodal covariates into the mark distribution would enable simultaneous inference on homophily and topological structure.

Finally, because HawkesNet explicitly couples timing and topology, future diagnostic tools must evolve to target this co-evolution directly; such as testing if temporal bursts coincide with motif formation at the expected pace, rather than relying solely on marginal temporal residuals or static network statistics. Scaling likelihood evaluation for massive networks also presents a significant future computational challenge. We provide an open-source implementation (\texttt{hawkesNet}) to facilitate these extensions and support reproducible research.

\section*{Acknowledgments}\label{sec:acks}

\if0\anon
{
 In preparing this manuscript, AI has been used in ways consistent with the Taylor \& Francis Author Use of AI Tools Policy, full responsibility for all ideas, arguments, and conclusions rests solely with the authors.
} \fi

\if1\anon
{

This work was supported by the Marsden Fund under grants 25-UOO-086 and 21-UOA-058 (UOA 3723517). The authors wish to acknowledge the use of New Zealand eScience Infrastructure (NeSI) high performance computing facilities as part of this research. New Zealand's national facilities are provided by NeSI and funded jointly by NeSI's collaborator institutions and through the Ministry of Business, Innovation \& Employment's Research Infrastructure program. In preparing this manuscript, AI has been used in ways consistent with the Taylor \& Francis Author Use of AI Tools Policy, full responsibility for all ideas, arguments, and conclusions rests solely with the authors.
} \fi

\spacingset{1}
\bibliography{bib.bib}
\newpage
\appendix

\section{Well-posedness of HawkesNet}
\label{sec:stability_theory}

This appendix develops the stability framework in which the ground rate is allowed to depend on the full marked network history. The technical route is the Poisson embedding and contraction method of \citep{bremaud1996stability}, adapted to a marked state space.

\subsection{Marked point processes via Poisson embedding}

Let $(\mathcal M,\mathcal P(\mathcal M))$ be a countable mark space equipped with the counting measure $\nu$.
Let $\Pi$ be a Poisson random measure (PRM) on
\[
E := \mathbb R \times \mathbb R_+ \times \mathcal M,
\qquad
\text{with intensity measure } \; dt \, dz \, \nu(dm).
\]
We write $\Pi(dt,dz,dm)$ for the associated random measure. We use the Campbell identity for PRMs: if $f\ge 0$ is measurable then
\[
\mathbb E\!\left[\int_E f(t,z,m)\,\Pi(dt,dz,dm)\right]
=
\int_E f(t,z,m)\,dt\,dz\,\nu(dm),
\]
see \citep[Ch. 6]{daley2003introduction}.

A (simple) marked point process $N$ is identified with its integer-valued random measure on $\mathbb R\times\mathcal M$.
For $t\in\mathbb R$, let $\mathcal H_t$ be the natural history $\sigma$-field generated by $N$ up to time $t$.
A predictable marked intensity is a family $\{\lambda(t,m)\}_{t\in\mathbb R,m\in\mathcal M}$ such that
$t\mapsto \lambda(t,m)$ is $\mathcal H_{t-}$-predictable for every $m$, $\lambda(t,m)\ge 0$, and
\[
\lambda_g(t) := \sum_{m\in\mathcal M}\lambda(t,m) < \infty
\quad\text{a.s. for a.e. }t.
\]

\subsection{Thinning equation}

Given a predictable intensity $\lambda$, define a marked point process $N^\lambda$ by
\begin{equation}
\label{eq:thinning_general}
N^\lambda(dt,dm)
:=
\int_{\mathbb R_+}\mathbf 1\{z\le \lambda(t,m)\}\,\Pi(dt,dz,dm).
\end{equation}
The stochastic integral in \eqref{eq:thinning_general} is in the standard sense for PRMs; by construction $N^\lambda$ is a simple marked point process
(\citep[Ch.~10]{daley2003introduction}; \citep{ogata1981lewis}).

\subsection{Total variation and symmetric difference}

For two (integer-valued) measures $\eta,\eta'$ on $\mathbb R\times\mathcal M$, let $|\eta-\eta'|$ denote the total variation measure of the signed measure $\eta-\eta'$. For integer-valued measures, $|\eta-\eta'|(A)$ equals the number of atoms that belong to exactly one of $\eta$ or $\eta'$ inside $A$ (i.e.\ the symmetric difference count).

Fix the weight function $w:\mathcal M\to[1,\infty)$ from Assumption~\ref{ass:HL}. For a signed measure $\mu$ on $\mathbb R\times\mathcal M$, define
\[
|\mu|_w(ds,dm):=w(m)\,|\mu|(ds,dm).
\]
In particular, for integer-valued measures $\eta,\eta'$,
\[
|\eta-\eta'|_w(A):=\int_A w(m)\,|\eta-\eta'|(ds,dm).
\]
For integer-valued measures, $|\eta-\eta'|_w(A)$ equals the sum of weights $w(m)$ over atoms in the symmetric difference of $\eta$ and $\eta'$ inside $A$.

\subsection{A coupling inequality}

The key estimate is the following coupling bound, which is the marked analogue of the standard thinning-coupling calculation
(see \citep{bremaud1996stability} and the PRM computations in \cite[Ch.~10]{daley2003introduction}).

\begin{lemma}[Coupling bound for marked thinnings]
\label{lem:coupling_marked}
Fix $T>0$ and $B\subseteq\mathcal M$.
Let $\lambda$ and $\lambda'$ be two predictable marked intensities such that
\(
\mathbb E\int_0^T\sum_{m\in B} w(m)\,(\lambda(t,m)+\lambda'(t,m))\,dt <\infty.
\)
Construct $N:=N^\lambda$ and $N':=N^{\lambda'}$ from the same PRM $\Pi$ via \eqref{eq:thinning_general}.
Then
\begin{equation}
\label{eq:coupling_bound_marked}
\mathbb E\bigl[\,|N-N'|_w([0,T]\times B)\,\bigr]
\le
\mathbb E\!\left[\int_0^T\sum_{m\in B} w(m)\,\bigl|\lambda(t,m)-\lambda'(t,m)\bigr|\,dt\right],
\end{equation}
where $|N-N'|_w(A):=\int_A w(m)\,|N-N'|(dt,dm)$.
\end{lemma}

\begin{proof}
By definition of $N$ and $N'$,
\[
|N-N'|_w([0,T]\times B)
=
\int_{[0,T]\times\mathbb R_+\times B}
w(m)\left|\mathbf 1\{z\le \lambda(t,m)\}-\mathbf 1\{z\le \lambda'(t,m)\}\right|
\,\Pi(dt,dz,dm).
\]
As in the unweighted proof, apply the PRM compensation formula and integrate over $z$:
for each fixed $(t,m)$,
\[
\int_0^\infty
\left|\mathbf 1\{z\le \lambda(t,m)\}-\mathbf 1\{z\le \lambda'(t,m)\}\right|\,dz
=
\bigl|\lambda(t,m)-\lambda'(t,m)\bigr|.
\]
Multiplying by $w(m)$ and using Tonelli yields \eqref{eq:coupling_bound_marked}.
\end{proof}

\subsection{HawkesNet as a fixed point in marked intensity form}

In HawkesNet we factor the marked intensity as
\begin{equation}
\label{eq:lambda_factor}
\lambda(t,m\mid \mathcal H_t)
=
\lambda_g(t\mid \mathcal H_t)\; q(m\mid t,\mathcal H_t),
\qquad
\sum_{m\in\mathcal M} q(m\mid t,\mathcal H_t)=1,
\end{equation}
but we do not assume $\lambda_g$ depends only on an unmarked history.
Both $\lambda_g$ and $q$ may depend on the full marked network history, and all stability assumptions are imposed on the full marked intensity $\lambda(\cdot,\cdot\mid\cdot)$.

We define the HawkesNet process $N$ as a fixed point of the thinning equation:
\begin{equation}
\label{eq:hawkesnet_fixed_point}
N(dt,dm)
=
\int_{\mathbb R_+} \mathbf 1\{z\le \lambda(t,m\mid \mathcal H_t)\}\,\Pi(dt,dz,dm).
\end{equation}

\subsection{Standing assumptions}
\label{sec:standing_assumptions}
Throughout this appendix we work under Assumptions~\ref{ass:H0}--\ref{ass:HL} from the main text.

\subsection{Existence, uniqueness, and non-explosion}

\begin{theorem}[Existence/uniqueness and non-explosion on $[0,\infty)$]
\label{thm:existence_unique_nonexplosion_marked}
Suppose Assumptions~\ref{ass:H0}--\ref{ass:HL} hold.
Then there exists a (pathwise) unique marked point process $N$ on $[0,\infty)\times\mathcal M$
adapted to the filtration of $\Pi$ that satisfies the fixed point equation \eqref{eq:hawkesnet_fixed_point} for all $t\ge 0$.
Moreover, for every $T>0$,
\[
\mathbb E\bigl[N([0,T]\times\mathcal M)\bigr] < \infty
\qquad\text{and hence}\qquad
N([0,T]\times\mathcal M) < \infty \ \text{a.s.}
\]
\end{theorem}

\begin{proof}
We construct the process by Picard iteration, starting from the null measure and repeatedly thinning the same driving PRM with the current iterate’s intensity. Define $N^{(0)}\equiv 0$ on $[0,\infty)\times\mathcal M$.
For $n\ge 0$, set
\[
\lambda^{(n)}(t,m) := \lambda(t,m\mid \mathcal H^{(n)}_t),
\quad t\ge 0,
\]
where $\mathcal H^{(n)}_t$ is the history generated by $N^{(n)}$ up to $t$.
Define $N^{(n+1)}$ from the same PRM $\Pi$ by thinning:
\begin{equation}
\label{eq:picard_marked}
N^{(n+1)}(dt,dm)
:=
\int_{\mathbb R_+}\mathbf 1\{z\le \lambda^{(n)}(t,m)\}\,\Pi(dt,dz,dm),
\qquad t\ge 0.
\end{equation}

We next show that successive iterates contract in expected total variation on $[0,T]\times \mathcal{M}$. Fix $T>0$ and define the weighted distance
\[
D_n^{(w)}(T) := \mathbb E\bigl[\,|N^{(n+1)}-N^{(n)}|_w([0,T]\times\mathcal M)\,\bigr].
\]
Note first that by Assumption~\ref{ass:H0},
\[
D_0^{(w)}(T)
=
\mathbb E\bigl[|N^{(1)}-N^{(0)}|_w([0,T]\times\mathcal M)\bigr]
=
\mathbb E\bigl[|N^{(1)}|_w([0,T]\times\mathcal M)\bigr]
=
\int_0^T \lambda_{g,\emptyset}^{(w)}(t)\,dt
\le
T\,\bar\lambda_\emptyset^{(w)}
<
\infty.
\]
We next verify the integrability condition required to apply Lemma~\ref{lem:coupling_marked}
to each Picard pair. Fix $T>0$. Applying Assumption~\ref{ass:HL} with $\eta=N^{(n)}$ and $\eta'=\mathbf 0$
gives, for each $t\in[0,T]$,
\[
\sum_{m\in\mathcal M} w(m)\lambda^{(n)}(t,m)
\le
\lambda_{g,\emptyset}^{(w)}(t)
+
\int_{[0,t)\times\mathcal M} h(t-s)\,|N^{(n)}|_w(ds,dm),
\]
since $\sum_m w(m)\lambda^{(n)}(t,m)
\le \sum_m w(m)\lambda_{\emptyset}(t,m)
+\sum_m w(m)\bigl|\lambda^{(n)}(t,m)-\lambda_{\emptyset}(t,m)\bigr|$.
Integrating over $t\in[0,T]$, taking expectations, and using Tonelli yields the recursion
\begin{align*}
\mathbb E\bigl[|N^{(n+1)}|_w([0,T]\times\mathcal M)\bigr]
&=
\int_0^T \mathbb E\!\left[\sum_{m\in\mathcal M} w(m)\lambda^{(n)}(t,m)\right]dt\\
&\le
\int_0^T \lambda_{g,\emptyset}^{(w)}(t)\,dt
+
\|h\|_1\,\mathbb E\bigl[|N^{(n)}|_w([0,T]\times\mathcal M)\bigr].
\end{align*}
Since $\mathbb E[|N^{(1)}|_w([0,T]\times\mathcal M)]=\int_0^T \lambda_{g,\emptyset}^{(w)}(t)dt<\infty$,
induction implies $\mathbb E[|N^{(n)}|_w([0,T]\times\mathcal M)]<\infty$ for all $n$.
Therefore the integrability condition in Lemma~\ref{lem:coupling_marked} holds for each Picard pair
$(\lambda^{(n)},\lambda^{(n-1)})$ on $[0,T]$.

Apply Lemma~\ref{lem:coupling_marked} with $B=\mathcal M$ to the pair $(N^{(n+1)},N^{(n)})$:
\begin{equation}
\label{eq:Dn_step1}
D_n^{(w)}(T)
\le
\mathbb E\!\left[\int_0^T\sum_{m\in\mathcal M} w(m)\bigl|\lambda^{(n)}(t,m)-\lambda^{(n-1)}(t,m)\bigr|\,dt\right].
\end{equation}
Now apply Assumption~\ref{ass:HL} with $\eta=N^{(n)}$ and $\eta'=N^{(n-1)}$.
For each $t\in[0,T]$,
\[
\sum_{m\in\mathcal M} w(m)\bigl|\lambda^{(n)}(t,m)-\lambda^{(n-1)}(t,m)\bigr|
\le
\int_{[0,t)\times\mathcal M} h(t-s)\,|N^{(n)}-N^{(n-1)}|_w(ds,dm).
\]
Insert this into \eqref{eq:Dn_step1} and use Tonelli to swap integrals:
\begin{align*}
D_n^{(w)}(T)
&\le
\mathbb E\!\left[
\int_0^T\int_{[0,t)\times\mathcal M} h(t-s)\,|N^{(n)}-N^{(n-1)}|_w(ds,dm)\,dt
\right]\\
&=
\mathbb E\!\left[
\int_{[0,T)\times\mathcal M}
\left(\int_s^T h(t-s)\,dt\right)
\,|N^{(n)}-N^{(n-1)}|_w(ds,dm)
\right]\\
&\le
\|h\|_1\,
\mathbb E\bigl[\,|N^{(n)}-N^{(n-1)}|_w([0,T)\times\mathcal M)\,\bigr]
=
\|h\|_1\,D_{n-1}^{(w)}(T).
\end{align*}
Iterating yields
\[
D_n^{(w)}(T) \le (\|h\|_1)^n D_0^{(w)}(T),\qquad n\ge0
\]
so \(\sum_{n\ge 0} D_n^{(w)}(T)<\infty\) because \(\|h\|_1<1\).

We now show almost sure convergence on $[0,T]$. Since $w\ge 1$,
\[
|N^{(n+1)}-N^{(n)}|([0,T]\times\mathcal M)\le |N^{(n+1)}-N^{(n)}|_w([0,T]\times\mathcal M),
\]
and the left-hand side is $\mathbb N$-valued. Markov's inequality yields
\[
\mathbb P\!\left(|N^{(n+1)}-N^{(n)}|([0,T]\times\mathcal M)\ge 1\right)
\le
\mathbb E\bigl[|N^{(n+1)}-N^{(n)}|([0,T]\times\mathcal M)\bigr]
\le
D_n^{(w)}(T).
\]
Since $\sum_{n\ge 0}D_n^{(w)}(T)<\infty$, the Borel--Cantelli lemma implies that
\[
|N^{(n+1)}-N^{(n)}|([0,T]\times\mathcal M)=0
\quad\text{for all sufficiently large $n$, a.s.}
\]
Hence there exists an (a.s. unique) integer-valued random measure $N^{(\infty,T)}$ on $[0,T]\times\mathcal M$
such that $N^{(n)}|_{[0,T]\times\mathcal M}=N^{(\infty,T)}$ for all large $n$, a.s.
Define $N|_{[0,T]}:=N^{(\infty,T)}$.

We proceed to pathwise identification of the fixed point on $[0,T]$. By the Borel--Cantelli argument above, there exists an a.s.\ finite random index $n_T$ such that
\[
|N^{(n+1)}-N^{(n)}|([0,T]\times\mathcal M)=0
\qquad\text{for all } n\ge n_T.
\]
In particular, for any $n\ge n_T$ we have
\[
N^{(n)}\big|_{[0,T]\times\mathcal M}
=
N^{(n+1)}\big|_{[0,T]\times\mathcal M}
=
N\big|_{[0,T]\times\mathcal M},
\]
so for every $t\le T$ the generated histories coincide: $\mathcal H^{(n)}_t=\mathcal H_t$.
Therefore, for all $t\le T$ and all $m\in\mathcal M$,
\[
\lambda^{(n)}(t,m)
=
\lambda(t,m\mid \mathcal H^{(n)}_t)
=
\lambda(t,m\mid \mathcal H_t).
\]
Restricting the Picard update \eqref{eq:picard_marked} to $[0,T]\times\mathcal M$ and using $N^{(n+1)}=N$ on $[0,T]$ gives
\[
N(dt,dm)
=
\int_{\mathbb R_+}\mathbf 1\{z\le \lambda(t,m\mid\mathcal H_t)\}\,\Pi(dt,dz,dm),
\qquad t\le T,
\]
so $N$ satisfies the fixed point equation \eqref{eq:hawkesnet_fixed_point} on $[0,T]$. Since $T>0$ was arbitrary, \eqref{eq:hawkesnet_fixed_point} holds for all $t\ge 0$.

We next prove non-explosion by bounding the expected total intensity $\mathbb E[\lambda_g(t)]$ and then integrating over $[0,T]$.
Define the weighted ground intensity of the constructed process by
\[
\lambda_g^{(w)}(t):=\sum_{m\in\mathcal M} w(m)\,\lambda(t,m\mid\mathcal H_t),
\qquad
\lambda_{g,\emptyset}^{(w)}(t):=\sum_{m\in\mathcal M} w(m)\,\lambda_{\emptyset}(t,m).
\]
Apply \eqref{eq:HL_lambda_marked} with $\eta=N$ and $\eta'=\mathbf 0$:
\[
\sum_{m\in\mathcal M} w(m)\bigl|\lambda(t,m\mid\mathcal H_t)-\lambda_\emptyset(t,m)\bigr|
\le
\int_{[0,t)\times\mathcal M} h(t-s)\,|N|_w(ds,dm).
\]
Since \(\lambda(t,m\mid\mathcal H_t)\ge 0\) and \(\lambda_\emptyset(t,m)\ge 0\),
\begin{align*}
\lambda_g^{(w)}(t)
=&
\sum_{m\in\mathcal M} w(m)\,\lambda(t,m\mid\mathcal H_t)\\
\le&
\sum_{m\in\mathcal M} w(m)\lambda_\emptyset(t,m)
+
\int_{[0,t)\times\mathcal M} h(t-s)\,|N|_w(ds,dm)\\
=&
\lambda_{g,\emptyset}^{(w)}(t)
+
\int_{[0,t)\times\mathcal M} h(t-s)\,|N|_w(ds,dm).
\end{align*}
Take expectations and use the compensator identity:
\[
\mathbb E\!\left[\int_{[0,t)\times\mathcal M} h(t-s)\,|N|_w(ds,dm)\right]
=
\int_0^t h(t-s)\,\mathbb E[\lambda_g^{(w)}(s)]\,ds.
\]
Therefore
\[
\mathbb E[\lambda_g^{(w)}(t)]
\le
\bar\lambda_\emptyset^{(w)}
+
\int_0^t h(t-s)\,\mathbb E[\lambda_g^{(w)}(s)]\,ds.
\]
Let $M^{(w)}(t):=\sup_{u\in[0,t]}\mathbb E[\lambda_g^{(w)}(u)]$. Then
\[
M^{(w)}(t)\le \bar\lambda_\emptyset^{(w)}+\|h\|_1 M^{(w)}(t),
\quad\text{so}\quad
M^{(w)}(t)\le \frac{\bar\lambda_\emptyset^{(w)}}{1-\|h\|_1}.
\]
Since
\[
\lambda_g(t):=\sum_{m\in\mathcal M}\lambda(t,m\mid\mathcal H_t)\le \lambda_g^{(w)}(t)
\]
(because \(w\ge1\)), for any \(T>0\),
\[
\mathbb E\bigl[N([0,T]\times\mathcal M)\bigr]
=
\int_0^T \mathbb E[\lambda_g(t)]\,dt
\le
\int_0^T \mathbb E[\lambda_g^{(w)}(t)]\,dt
\le
T\cdot \frac{\bar\lambda_\emptyset^{(w)}}{1-\|h\|_1}
<\infty.
\]
Thus $N([0,T]\times\mathcal M)<\infty$ a.s.

Finally, we establish pathwise uniqueness. Let $N$ and $N'$ be two solutions of \eqref{eq:hawkesnet_fixed_point} driven by the same PRM $\Pi$ (on the same probability space). Fix $T>0$ and define
\[
D^{(w)}(T):=\mathbb E\bigl[|N-N'|_w([0,T]\times\mathcal M)\bigr].
\]
By the non-explosion estimate proved above (applied to the weight $w$),
\[
\mathbb E\bigl[|N|_w([0,T]\times\mathcal M)\bigr]
=
\int_0^T \mathbb E[\lambda_g^{(w)}(t)]\,dt
\le
T\cdot\frac{\bar\lambda_\emptyset^{(w)}}{1-\|h\|_1}
<\infty,
\]
and similarly for $N'$, hence $D^{(w)}(T)<\infty$.

Applying Lemma~\ref{lem:coupling_marked} with $B=\mathcal M$ to the pair of intensities
$\lambda(t,m\mid\mathcal H_t)$ and $\lambda(t,m\mid\mathcal H'_t)$ yields
\begin{align*}
D^{(w)}(T)
&\le
\mathbb E\!\left[\int_0^T\sum_{m\in\mathcal M} w(m)\,
\bigl|\lambda(t,m\mid \mathcal H_t)-\lambda(t,m\mid \mathcal H'_t)\bigr|\,dt\right]\\
&\le
\mathbb E\!\left[\int_0^T\int_{[0,t)\times\mathcal M} h(t-s)\,|N-N'|_w(ds,dm)\,dt\right]\\
&=
\mathbb E\!\left[\int_{[0,T)\times\mathcal M}\left(\int_s^T h(t-s)\,dt\right)\,|N-N'|_w(ds,dm)\right]\\
&\le
\|h\|_1\,\mathbb E\bigl[|N-N'|_w([0,T)\times\mathcal M)\bigr]
=
\|h\|_1\,D^{(w)}(T).
\end{align*}
Since $\|h\|_1<1$, we conclude $D^{(w)}(T)=0$, hence $|N-N'|_w([0,T)\times\mathcal M)=0$ almost surely.
Because $w\ge 1$, this implies $|N-N'|([0,T)\times\mathcal M)=0$ almost surely, i.e.\ $N=N'$ on $[0,T)\times\mathcal M$.
As $T>0$ was arbitrary, $N=N'$ almost surely on $[0,\infty)\times\mathcal M$.
\end{proof}

\subsection{Proofs of auxiliary results from Section~\ref{sec:path_dependent}}
\label{sec:aux_proofs_path_dependent}

\begin{proof}[Proof of Lemma~\ref{lem:HL_no_uniform_envelope}]
Fix \(t\ge 0\) and histories \(\eta,\eta'\). Write
\[
\lambda(t,m\mid\eta_t)=\lambda_g(t\mid\eta_t)\,q(m\mid t,\eta_t),
\qquad
\lambda(t,m\mid\eta'_t)=\lambda_g(t\mid\eta'_t)\,q(m\mid t,\eta'_t).
\]
Using the shorthand
\[
\lambda_g:=\lambda_g(t\mid\eta_t),\qquad
\lambda_g':=\lambda_g(t\mid\eta'_t),\qquad
q:=q(\cdot\mid t,\eta_t),\qquad
q':=q(\cdot\mid t,\eta'_t),
\]
we have
\begin{align*}
\sum_{m\in\mathcal M} w(m)\,\bigl|\lambda_g q(m)-\lambda_g' q'(m)\bigr|
&\le
\sum_{m\in\mathcal M} w(m)\,\bigl|(\lambda_g-\lambda_g')q(m)\bigr|
+
\sum_{m\in\mathcal M} w(m)\,\lambda_g'\,\bigl|q(m)-q'(m)\bigr| \\
&=
|\lambda_g-\lambda_g'| \sum_{m\in\mathcal M} w(m)\,q(m)
+
\lambda_g' \sum_{m\in\mathcal M} w(m)\,\bigl|q(m)-q'(m)\bigr| \\
&\le
\bar\mu_w\,|\lambda_g-\lambda_g'|
+
\sum_{m\in\mathcal M}\lambda_g(t\mid\eta'_t)\,w(m)\,\bigl|q(m\mid t,\eta_t)-q(m\mid t,\eta'_t)\bigr|.
\end{align*}
Applying the two assumed bounds and collecting terms gives
\[
\sum_{m\in\mathcal M} w(m)\,
\bigl|\lambda(t,m\mid\eta_t)-\lambda(t,m\mid\eta'_t)\bigr|
\le
\int_{[0,t)\times\mathcal M}
\bigl(\bar\mu_w h_g+h_{q,\lambda}\bigr)(t-s)\,|\eta-\eta'|_w(ds,dm),
\]
which is Assumption~\ref{ass:HL} with \(h:=\bar\mu_w h_g+h_{q,\lambda}\).
\end{proof}

\begin{proof}[Proof of Lemma~\ref{lem:alpha_implies_HL}]
Fix \(t\ge 0\) and histories \(\eta,\eta'\). Since \(\eta_t\) and \(\eta'_t\) are finite integer-valued measures on
\([0,t)\times\mathcal M\), there exist finitely many unit atoms
\(\{(s_k,m_k)\}_{k=1}^K \subset [0,t)\times\mathcal M\) (allowing repetitions if a location has multiplicity difference greater than one)
such that \(\eta'_t\) can be obtained from \(\eta_t\) by toggling these atoms one at a time. Let
\[
\eta^{(0)}:=\eta_t,
\qquad
\eta^{(k)}:=\eta^{(k-1)} \text{ with the atom } (s_k,m_k) \text{ toggled},
\qquad
k=1,\dots,K,
\]
so that \(\eta^{(K)}=\eta'_t\).

By the triangle inequality,
\[
\sum_{m\in\mathcal M}
w(m)\bigl|\lambda(t,m\mid\eta_t)-\lambda(t,m\mid\eta'_t)\bigr|
\le
\sum_{k=1}^K
\sum_{m\in\mathcal M}
w(m)\bigl|\lambda(t,m\mid\eta^{(k-1)})-\lambda(t,m\mid\eta^{(k)})\bigr|.
\]
For each \(k\), the pair \((\eta^{(k-1)},\eta^{(k)})\) differs by exactly one unit atom \((s_k,m_k)\), so by Definition~\ref{def:one_event_influence},
\[
\sum_{m\in\mathcal M}
w(m)\bigl|\lambda(t,m\mid\eta^{(k-1)})-\lambda(t,m\mid\eta^{(k)})\bigr|
\le
w(m_k)\,\alpha_w(t-s_k).
\]
Summing over \(k\) yields
\[
\sum_{m\in\mathcal M}
w(m)\bigl|\lambda(t,m\mid\eta_t)-\lambda(t,m\mid\eta'_t)\bigr|
\le
\int_{[0,t)\times\mathcal M}\alpha_w(t-s)\,|\eta-\eta'|_w(ds,dm),
\]
which is \eqref{eq:HL_from_alpha}.

For the converse bound, assume Assumption~\ref{ass:HL} holds with envelope \(h\). Fix \(u>0\), \(t\ge u\), a history \(\eta\), and a mark \(m_0\in\mathcal M\). Apply \eqref{eq:HL_lambda_marked} to the pair
\(\eta_t+\delta_{(t-u,m_0)}\) and \(\eta_t\):
\[
\sum_{m\in\mathcal M}
w(m)\Bigl|
\lambda\!\left(t,m \mid \eta_t+\delta_{(t-u,m_0)}\right)
-
\lambda\!\left(t,m \mid \eta_t\right)
\Bigr|
\le
w(m_0)\,h(u).
\]
Divide by \(w(m_0)\) and take the supremum over \(t,\eta,m_0\). This gives
\[
\alpha_w(u)\le h(u),\qquad u>0.
\]
Hence \(\alpha_w\) is pointwise minimal among all envelopes satisfying \eqref{eq:HL_lambda_marked}.
\end{proof}

\begin{proof}[Proof of Corollary~\ref{cor:alpha_wellposed}]
By Lemma~\ref{lem:alpha_implies_HL}, Assumption~\ref{ass:HL} holds with \(h=\alpha_w\). Together with Assumption~\ref{ass:H0}, the hypotheses of Theorem~\ref{thm:existence_unique_nonexplosion_marked} are satisfied.
\end{proof}

\begin{proof}[Proof of Corollary~\ref{cor:linear_growth}]
The proof of Theorem~\ref{thm:existence_unique_nonexplosion_marked} established the uniform bound
\[
\sup_{t\ge 0}\mathbb E[\lambda_g(t)]
\le
\sup_{t\ge 0}\mathbb E[\lambda_g^{(w)}(t)]
\le
C_\lambda
:=
\frac{\bar\lambda_\emptyset^{(w)}}{1-\|h\|_1}.
\]
Hence, for every \(T>0\),
\[
\mathbb E\bigl[N_g([0,T])\bigr]
=
\int_0^T \mathbb E[\lambda_g(t)]\,dt
\le
C_\lambda\,T.
\]
By Assumption~\ref{ass:bounded_update},
\[
N_T^{\text{nodes}}
\le
N_0 + B_V\,N_g([0,T]),
\qquad
|E_T|
\le
|E_0| + B_E\,N_g([0,T]),
\]
and taking expectations yields the first two claimed bounds.

For the aged edge mass, decompose \(E_t\) into the initial edges and the edges born after time \(0\). Since each update adds at most \(B_E\) new edges,
\[
|E_t|^{(a)}
\le
|E_0|\,\|a\|_\infty
+
B_E\int_{[0,t)\times\mathcal M} a(t-s)\,N(ds,dm).
\]
Taking expectations and using the compensator identity gives
\begin{align*}
\mathbb E[|E_t|^{(a)}]
&\le
|E_0|\,\|a\|_\infty
+
B_E\int_0^t a(t-s)\,\mathbb E[\lambda_g(s)]\,ds \\
&\le
|E_0|\,\|a\|_\infty
+
B_E\,C_\lambda \int_0^\infty a(u)\,du \\
&=
|E_0|\,\|a\|_\infty + B_E\,C_\lambda\,\|a\|_1.
\end{align*}
The right-hand side does not depend on \(t\), so \(\sup_{t\ge 0}\mathbb E[|E_t|^{(a)}] < \infty\).
\end{proof}

\subsection{Induced HawkesNet models: stability by predictable marking}
\label{sec:induced_marking_wellposed}

\begin{proposition}[Predictable marking preserves well-posedness and non-explosion]
\label{prop:induced_marking_wellposed}
Let $N_g$ be a simple point process on $\mathbb R_+$ with natural filtration $\{\mathcal H_t^g\}_{t\ge 0}$ and
$\mathcal H_{t-}^g$-predictable intensity $\lambda_g(t\mid \mathcal H_t^g)$ (with respect to $dt$).
Assume that for every $T>0$, $N_g([0,T])<\infty$ almost surely ($\mathbb E[N_g([0,T])]<\infty$).

Let $\mathcal M$ be a countable mark space and let $q(\cdot\mid t,\mathcal H_t)$ be an $\mathcal H_{t-}$-predictable PMF on $\mathcal M$
with $q(m\mid t,\mathcal H_t)=0$ for $m\notin\mathcal M_t$.
Construct a marked point process
\[
N=\sum_{i\ge 1}\delta_{(t_i,m_i)}
\]
by attaching to each event time $t_i$ of $N_g$ a mark $m_i$ sampled from $q(\cdot\mid t_i,\mathcal H_{t_i})$
(using additional randomness independent of $\mathcal H_{t_i}$).

Then $N$ is a simple marked point process with conditional intensity
\[
\lambda(t,m\mid \mathcal H_t)=\lambda_g(t\mid \mathcal H_t^g)\,q(m\mid t,\mathcal H_t),
\]
and it is non-explosive on finite horizons:
\[
N([0,T]\times\mathcal M)=N_g([0,T])\quad\text{for all }T>0.
\]
\end{proposition}

\begin{proof}
Fix an enumeration \(\mathcal M=\{m^{(1)},m^{(2)},\ldots\}\), let \(0<t_1<t_2<\cdots\) be the event times of \(N_g\), and let
\((U_i)_{i\ge1}\) be i.i.d.\ \(\mathrm{Uniform}(0,1)\) random variables independent of \(N_g\).

We construct the marks recursively. Suppose \(M_1,\ldots,M_{i-1}\) have already been defined, and let \(\mathcal H_{t_i}\) denote the marked history generated by
\(\mathcal H_{t_i}^g\) together with the past marked events \(\{(t_j,M_j):j<i\}\). Define
\[
F_i(k):=\sum_{\ell=1}^k q\bigl(m^{(\ell)}\mid t_i,\mathcal H_{t_i}\bigr),
\qquad
F_i(0):=0,
\]
and set \(M_i=m^{(k)}\) whenever \(F_i(k-1)<U_i\le F_i(k)\). Since \(q(\cdot\mid t_i,\mathcal H_{t_i})\) is a PMF on the countable space \(\mathcal M\), this defines \(M_i\) uniquely and
\[
\mathbb P(M_i=m\mid \mathcal H_{t_i})=q(m\mid t_i,\mathcal H_{t_i}),
\qquad m\in\mathcal M.
\]
Because \(q(m\mid t_i,\mathcal H_{t_i})=0\) for \(m\notin\mathcal M_{t_i}\), the mark is admissible. Define
\[
N:=\sum_{i\ge1}\delta_{(t_i,M_i)}.
\]

The event times of \(N\) are exactly those of \(N_g\). Since \(N_g\) is simple, \(N\) is simple, and for every \(T>0\),
\[
N([0,T]\times\mathcal M)=N_g([0,T])<\infty
\qquad\text{a.s.}
\]
Hence \(N\) is non-explosive on finite horizons.

To identify the marked intensity, enlarge the ground-process filtration by the independent uniforms and set
\[
\widetilde{\mathcal H}_t:=\sigma\!\bigl(\mathcal H_t^g,\; U_i\mathbf 1_{\{t_i<t\}},\ i\ge1\bigr).
\]
Because the additional uniforms are independent of \(N_g\), the \(\widetilde{\mathcal H}_t\)-intensity of \(N_g\) is still
\(\lambda_g(t\mid\mathcal H_t^g)\). Moreover, each past mark \(M_i\) is a measurable function of \((t_j,U_j)_{j\le i}\), so the marked filtration \(\mathcal H_t\) is contained in \(\widetilde{\mathcal H}_t\).

Let \(f:\mathbb R_+\times\mathcal M\times\Omega\to[0,\infty)\) be \(\mathcal H_t\)-predictable. Then
\begin{align*}
\mathbb E\!\left[\int_{\mathbb R_+\times\mathcal M} f(t,m)\,N(dt,dm)\right]
&=
\mathbb E\!\left[\sum_{i\ge1} f(t_i,M_i)\right] \\
&=
\mathbb E\!\left[\sum_{i\ge1}\sum_{m\in\mathcal M} f(t_i,m)\,q(m\mid t_i,\mathcal H_{t_i})\right].
\end{align*}
The process
\[
g(t):=\sum_{m\in\mathcal M} f(t,m)\,q(m\mid t,\mathcal H_t)
\]
is nonnegative and \(\widetilde{\mathcal H}_t\)-predictable, so the compensator identity for \(N_g\) gives
\begin{align*}
\mathbb E\!\left[\sum_{i\ge1}\sum_{m\in\mathcal M} f(t_i,m)\,q(m\mid t_i,\mathcal H_{t_i})\right]
&=
\mathbb E\!\left[\int_0^\infty g(t)\,\lambda_g(t\mid\mathcal H_t^g)\,dt\right] \\
&=
\mathbb E\!\left[\int_0^\infty \sum_{m\in\mathcal M}
f(t,m)\,\lambda_g(t\mid\mathcal H_t^g)\,q(m\mid t,\mathcal H_t)\,dt\right].
\end{align*}
Since \(\nu\) is counting measure on \(\mathcal M\), this is
\[
\mathbb E\!\left[\int_0^\infty\int_{\mathcal M}
f(t,m)\,\lambda_g(t\mid\mathcal H_t^g)\,q(m\mid t,\mathcal H_t)\,\nu(dm)\,dt\right].
\]
Thus the \(\mathcal H_t\)-conditional intensity of \(N\) is
\[
\lambda(t,m\mid\mathcal H_t)=\lambda_g(t\mid\mathcal H_t^g)\,q(m\mid t,\mathcal H_t).
\]
\end{proof}

\begin{corollary}[BA and CS induced HawkesNet models are well-posed]
\label{cor:BA_CS_wellposed}
Consider the induced specification \eqref{eq:lambda_factor_induced} in which the ground process $N_g$ is non-explosive.
If $q$ is taken to be the BA mark mechanism of Example~\ref{ex:BA} or the CS mark mechanism of Example~\ref{ex:cs}
(with a finite candidate set $\mathcal C_t$ at each event time), then the resulting marked process exists and is non-explosive on every finite time window.
\end{corollary}

\begin{proof}
In both Examples~\ref{ex:BA} and~\ref{ex:cs}, the update rule at each event time is given by a finite sequence of sampling steps on a finite candidate set. Hence the rule induces an exact conditional PMF on the countable mark space \(\mathcal M\), supported on admissible marks. Proposition~\ref{prop:induced_marking_wellposed} therefore applies.

The product forms \eqref{eq:bernoulli} and \eqref{eq:csdensity} are computational approximations to these exact PMFs and are not needed for the well-posedness claim.
\end{proof}

\newpage

\section{Nodal Covariates and Joint Estimation}\label{app:gender}

Edges in a network may depend not only on network structure, but on nodal covariates. In this section we discuss a natural extension to the CS model that facilitates this.

The CS HawkesNet framework incorporates nodal covariates by factoring the mark density into structural and attribute components. At each event time $t$, we observe $N_t^{\text{new}} \sim \text{Poisson}(\lambda_{\text{nodes}})$ new nodes. Each new node $v \in \mathcal{V}_t^{\text{new}}$ is assigned a covariate value $x_v$ drawn from a distribution $f_X(\cdot\,;\xi)$ with parameters $\xi$. The joint mark density $q(m_t \mid t, \mathcal{H}_t)$ is the product of the structural formation probability and the attribute likelihood:
\[
q(m_t \mid t, \mathcal{H}_t) \;=\; \underbrace{\text{Poisson}(N_t^{\text{new}}; \lambda)}_{\text{Node Count}} \;\times\; \underbrace{\prod_{(i,j) \in \mathcal{C}_t} \bigl(p_{i,j}^{\text{CS}}\bigr)^{e_{i,j}} \bigl(1 - p_{i,j}^{\text{CS}}\bigr)^{1-e_{i,j}}}_{\text{Edge Formation}} \;\times\; \underbrace{\prod_{v \in \mathcal{V}_t^{\text{new}}} f_X(x_v; \xi)}_{\text{Attribute Likelihood}},
\]
where $\mathcal{C}_t$ is the set of candidate dyads at time $t$. For a categorical attribute with $L$ levels, we employ an $(L-1)$ multinomial parameterization where $\mathbb{P}(X = \ell) = \pi_\ell$ for $\ell \in \{1, \dots, L-1\}$ and the reference level probability is $1 - \sum \pi_\ell$.

This factorization enables the inclusion of homophily and mixing effects within the change-statistic vector $C_{i,j}$. By extending $C_{i,j}$ to include terms such as \texttt{nodeMatch} or \texttt{nodeMix}, the edge formation probability $p_{i,j}^{\text{CS}}$ becomes a function of both the network topology and the nodal attributes. The parameters $\xi$ and the structural parameters $\theta$ are estimated jointly via MLE, as the log-likelihood remains additive across the three components. This approach is consistent with the Exponential Random Network Model (ERNM) framework \citep{fellows2012, fellows2023ernm}, but adapted here for the growing, time-stamped context of HawkesNet.

\newpage
\section{Further Parametric examples}\label{app:stationarity}

\begin{remark}
    Including the triangle term in the change statistic model induces global dependence among all edges. This leads to related issues of near degeneracy \citep{handcock2003} where significant probability mass is placed on empty and complete networks. However, our inclusion of the time decay term mitigates this as the global dependence diminishes over time, a full investigation of degeneracy is beyond the scope of this work.
\end{remark}

\subsection{Nonlinear Hawkes example: ETAS model}\label{app:ETAS}

We justify mark path dependence by illustrating its appearance in the popular ETAS model \citep{ogata1998space}. If we specify the conditional intensity
\[
\lambda(t,m\mid\mathcal{H}_t)=\lambda_0+\kappa \sum_{i:t_i<t}\exp(\alpha(m_i-m_0))\left(1+\frac{t-t_i}{c}\right)^{-p},
\]
we have a linear marked Hawkes process. Now consider instead
\begin{equation}\label{eq:nonlinearETAS}
\lambda(t,m\mid\mathcal{H}_t)=\lambda_0+\kappa \sum_{i:t_i<t}\exp(\alpha(m_i-m_0)+\beta\max_{j<i} \{m_j\})\left(1+\frac{t-t_i}{c}\right)^{-p}.
\end{equation}
In \eqref{eq:nonlinearETAS}, each event’s contribution depends not only on its mark, but also on all previous marks. The kernel is a function of the entire history. Note that this mark-path dependence is weak in the sense that the candidate mark $m$ does not have a separate density: $\max_{j<i}m_j$ is fixed once event $i$ is born. To make \ref{eq:nonlinearETAS} strongly mark-path dependent, the extra factor must depend on the current history, i.e. some statistic $\mathcal{S}_i(t,\mathcal{H}_t)$ that is non constant in $t$.

\subsection{Additional induced models from network growth procedures}\label{app:more_induced}

\begin{example}[Stochastic block model HawkesNet]
Stochastic block models are a popular framework for modeling communities in networks \citep{block_models_1,block_models_2,block_models_3}. A block-model-inspired mark mechanism can be specified as follows. For an $r$-community model, let $c^\star\in\mathbb R^r$ be group membership probabilities and $C\in\mathbb R^{r\times r}$ be between-group edge probabilities. At each event:
(i) add a new node at time $t$; (ii) draw its community label $x_{N_t}\sim\mathrm{Multinomial}(c^\star)$; (iii) connect it to each existing node $i$ with probability
\[
p_i^{BM}=\exp(-\tau(t-t_i))\,C_{x_i,x_{N_t}}.
\]
This yields the product-Bernoulli PMF
\[
q(m\mid t,\mathcal H_t)=\prod_{i=1}^{N_{t-}} (p_i^{BM})^{e_i}(1-p_i^{BM}(t))^{1-e_i}.
\]
\end{example}

\begin{example}[Latent space HawkesNet]
Latent space models posit a latent social space with edge probabilities depending on distance \citep{latent_space_1,latent_space_2,latent_space_3}. For an $r$-dimensional latent space: (i) add a new node at time $t$ with latent position $x_{N_t}\in\mathbb R^r$; (ii) connect to each existing node $i$ with probability
\[
p_i^{LS}=\exp(-\tau(t-t_i))\cdot \frac{1}{1+\exp(-\theta^\top d_{i,N_t})},
\]
where $d_{i,N_t}$ is Euclidean distance between $x_i$ and $x_{N_t}$. This again induces a product-Bernoulli PMF over candidate edges.
\end{example}

\newpage

\section{Simulation Study Details}\label{app:sim_study}

In this section we demonstrate parameter recovery for the BA and CS induced kernels. We investigate finite-sample properties of the proposed models and the unbiasedness of maximum likelihood estimates.

Realizations of the marked point process are simulated as in Section~\ref{sec:simulation}: a complete history of network updates $\{(t_i, m_i)\}_{i=1}^n$ on $[0, T)$ for each realization, equivalently a time-evolving network $\mathcal{G}_T$ with time-stamped edge and node entries.

We simulate $100$ realizations per design, and estimate parameters by MLE as in Section~\ref{sec:likelihood}. Exponential temporal decay is used throughout. Optimization is performed with the Nelder-Mead algorithm; invalid parameter regions (e.g.\ $m \leq 0$) are penalized in the objective so that estimates remain in the valid domain.

\subsection{Preferential Attachment (BA) HawkesNet}\label{sec:BA_sim}

The BA-inspired HawkesNet is motivated by preferential attachment: new nodes are more likely to attach to high-degree nodes. \citet{BA_model} show that a model in which attachment probability is proportional to degree yields a power-law degree distribution $p(k) \propto k^{-\gamma}$ with $\gamma = 3$. In our framework we refer to this as the BA HawkesNet model. At each event, one new node is added and $K_t \sim \text{Poisson}(m)$ edges are drawn (with $m$ the expected number of edges per event); attachment targets are chosen without replacement with probabilities proportional to time-decayed degree (Example~\ref{ex:BA}).

Simulation parameters were chosen to give a mean degree of approximately $2$ and total nodes around $500$. The decay is moderate with $\tau = 1.0$, $\exp(-\tau \cdot 1) \approx 0.37$, so excitation one time unit after an event is about $37\%$ of its immediate value; with $\beta$ (Hawkes decay) equal to $1$, $\exp(-\beta \cdot 1) \approx 0.37$.

\begin{table}[!htb]
    \centering
    \begin{tabular}{lccc}
        \toprule
        Parameter & True & Mean estimate & Std.\ dev. \\
        \midrule
        $T$ & 25 & \texttt{--} & \texttt{--} \\
        \midrule
        $\lambda_{\emptyset}$ (baseline rate) & 10 & 12.10 & 3.26 \\
        $\beta$ (Hawkes decay) & 1.0 & 1.47 & 1.20 \\
        $K$ (triggering weight) & 0.5 & 0.46 & 0.37 \\
        \midrule
        $\tau$ (mark decay) & 1.0 & 1.00 & 0.05 \\
        $m$ (expected edges/event) & 1.0 & 1.00 & 0.05 \\
        \midrule
    \end{tabular}
    \caption{BA HawkesNet: parameters and summary of MLE over $100$ replications (simulate then fit). Top block: temporal point process; bottom block: mark kernel.}
    \label{tab:BA_table}
\end{table}

Table \ref{tab:BA_table} reports simulation results. MLE recovers the true parameters on average; the variability of $\beta$, $K$, $\lambda_{\emptyset}$ is consistent with the flat likelihood surfaces commonly observed in Hawkes models \citep{schoenberg_density_trick,kresin2023parametric}. Figure \ref{fig:BA_sim_study_shape} in Appendix \ref{sec:mle_dist} shows the distribution of the parameter estimates. Appendix~\ref{sec:app_consistency} shows that estimates become more stable as the time window $T$ increases (MLE consistency, \citealt{rathbun1996asymptotic}). We show results in Table \ref{tab:BA_table} rather than with parameters that produce larger node sets (as in Appendix~\ref{sec:app_consistency})  to emphasize the difficulty of fitting Hawkes model parameters.  Appendix~\ref{sec:app_explosivity_BA} illustrates a supercritical regime when the Hawkes memory decays slowly and explains why the non-explosivity condition is important for reliable MLE. 

\subsection{Change Statistic (CS) HawkesNet}\label{sec:CS_sim}

The CS HawkesNet (Example~\ref{ex:cs}) adds $N_t^{\text{new}} \sim \text{Poisson}(\lambda_{\text{nodes}})$ nodes at each event and draws the number of new edges $K_t \sim \text{Poisson}(m)$, selecting which edges to add via a logistic model on change statistics with time decay. Candidate edges are restricted to a truncated window of the 100 most recent nodes, keeping the candidate set $\mathcal{O}(n)$. Parameters were chosen to produce sparse networks with a strong transitivity effect (triangle coefficient $3.0$), and small positive 2-star and negative 3-star terms as commonly used in ERGM-style specifications \citep{Fellows2018,snijders2006}.

Recall from Section~\ref{sec:likelihood} that the log-likelihood decomposes as
\begin{align}\label{eq:CS_identifiability}
\log \mathcal{L}(\theta) &= \underbrace{\sum_{i=1}^{N} \log q(m_i \mid t_i, \mathcal{H}_{t_i})}_{\text{mark contribution}} \;+\; \underbrace{\sum_{i=1}^{N} \log\!\Bigl(\lambda_{\emptyset} + K\!\sum_{j<i} e^{-\beta(t_i - t_j)}\Bigr)}_{\text{ground intensity}} \; \\&-\; \underbrace{\Bigl(\lambda_{\emptyset}\, T + \frac{K}{\beta}\sum_{i=1}^{N}\bigl(1 - e^{-\beta(T-t_i)}\bigr)\Bigr)}_{\text{compensator}}.
\end{align}
Because $q$ is a PMF that integrates to one over $\mathcal{M}_t$, the compensator depends only on the temporal parameters $(\lambda_{\emptyset}, K, \beta)$ and not on any mark parameters. The mark contribution likewise depends only on $(\theta, \tau, \lambda_{\text{nodes}}, m)$ via Equation \eqref{eq:csdensity}. In the log-likelihood these two groups of parameters are therefore additively separable: the ground intensity and compensator terms involve $(\lambda_{\emptyset}, K, \beta)$ alone, while the mark sum involves $(\theta, \tau, \lambda_{\text{nodes}}, m)$ alone.

Despite this separability, practical identification issues can arise. We retain a slight overspecification of our model to demonstrate this point. Typically, change statistic models include an edge intercept term $\theta_1$ governs the baseline probability that any given candidate edge is formed. This role overlaps is however taken by $m$, which directly controls the expected number of edges per event via the $\text{Poisson}(m)$ draw. This renders the edge intercept unnecessary, we retain it, though fix it to retain the connection to ERGMs, LOLOGs and other change statistic models. This anchors the baseline edge scale and allows unbiased recovery of $m$ and the structural change statistics $(\theta_2, \theta_3, \theta_4)$.

\begin{table}[tb]
    \centering
    \begin{tabular}{lccc}
        \toprule
        Parameter & True & Mean estimate & Std.\ dev. \\
        \midrule
        $T$ & 50 & \texttt{--} & \texttt{--} \\
        \midrule
        $\lambda_{\emptyset}$ & 10.0 & 8.666 & 1.571 \\
        $K$ & 0.5 & 0.259 & 0.265 \\
        $\beta$ & 2.0 & 1.636 & 1.309 \\
        \midrule
        edges ($\theta_1$) & $-7.0$ & (fixed) & \texttt{--} \\
        triangles ($\theta_2$) & $3.0$ & 3.040 & 0.165 \\
        2-star ($\theta_3$) & $0.1$ & 0.445 & 0.065 \\
        3-star ($\theta_4$) & $-0.1$ & -0.256 & 0.041 \\
        $\tau$ & 1.0 & 0.715 & 0.036 \\
        $\lambda_{\text{nodes}}$ & 1.0 & (fixed) & \texttt{--} \\
        $m$ & 1.0 & 1.006 & 0.045 \\
        \midrule
    \end{tabular}
    \caption{CS HawkesNet: true parameters and summary of MLE over replications. The edge intercept $\theta_1$ and $\lambda_{\text{nodes}}$ are fixed; all other parameters are estimated. T = 50 produces around $500$ events}
    \label{tab:CS_table}
\end{table}

Table~\ref{tab:CS_table} summarizes parameter recovery for the CS model, for a specification producing around $500$ events. Appendix~\ref{sec:app_consistency} confirms that bias and variance decay as the observation window $T$ increases.

\subsection{Simulation algorithm}\label{sec:sim_algorithm}

Algorithm \ref{alg:ogata_thinning} show the algorithm of \cite{ogata1981lewis} applied to our marked point process with dynamic mark space.

\begin{algorithm}[!h]
\caption{Ogata's Thinning Algorithm for HawkesNet}
\label{alg:ogata_thinning}
\begin{algorithmic}[1]

\REQUIRE Dominating intensity function $\Lambda(t)$ such that
$\lambda_g(t \mid \mathcal{H}_t) \leq \Lambda(t)$ for all $t \leq T_{\max}$, and a method to sample event times from the inhomogeneous Poisson process with rate $\Lambda(t)$.

\ENSURE A list of accepted events (time $t_i$ and mark $m_i$) up to $T_{\max}$.

\STATE Initialize current time: $t \gets 0$, event list: $ N \gets \emptyset$.

\WHILE {$t < T_{\max}$}
    \STATE Sample a waiting time $\Delta \sim \mathrm{Exp}(\Lambda(t))$.
    \STATE Propose a candidate event time $t^* \gets t + \Delta$.
    \IF {$t^* > T_{\max}$}
        \STATE \textbf{break}
    \ENDIF
    \STATE Compute the true ground intensity at $t^*$: $\lambda^* \gets \lambda_g(t^* \mid \mathcal{H}_{t^*})$.
    \STATE Compute acceptance probability: $p \gets \lambda^* \,/\, \Lambda(t^*)$.
    \STATE Draw $U \sim \mathrm{Uniform}(0,1)$.
    \IF {$U \le p$}
        \STATE Sample $m^*$ from $q(m^*\mid t^*,\mathcal H_{t^*})=\lambda(t^*,m^*\mid\mathcal H_{t^*})/\lambda^*$.
        \STATE Append $(t^*,m^*)$ to the event list $ N$.
    \ENDIF
    \STATE $t \gets t^*$
\ENDWHILE

\STATE \textbf{return} $ N$

\end{algorithmic}
\end{algorithm}
\break

\subsection{Empirical consistency verification}\label{sec:app_consistency}

We evaluate the consistency of the Maximum Likelihood Estimators (MLE) by performing a series of simulate-and-fit experiments across increasing observation windows. This procedure assesses the asymptotic behavior of the estimators as the expected number of events $E[N(T)]$ grows. For each window, we generate $N_{\text{rep}} = 25$ independent realizations from the true model and compute the MLE for each.

For each parameter $\theta$, we compute estimate the root mean squared error (RMSE):
\[ \text{RMSE}_T = \sqrt{\frac{1}{N_{\text{rep}}} \sum_{i=1}^{N_{\text{rep}}} (\hat{\theta}_{i,T} - \theta_{\text{true}})^2} \]

As $T$ increases, we observe that the distribution of estimates centers more tightly around $\theta_{\text{true}}$ and the $\text{RMSE}_T$ decays toward zero, confirming that estimation error decreases as the expected number of events grows (an empirical large-sample trend).

\subsubsection{BA Model Consistency.} 

\begin{figure}[H]
    \centering
    \includegraphics[width=1.0\textwidth]{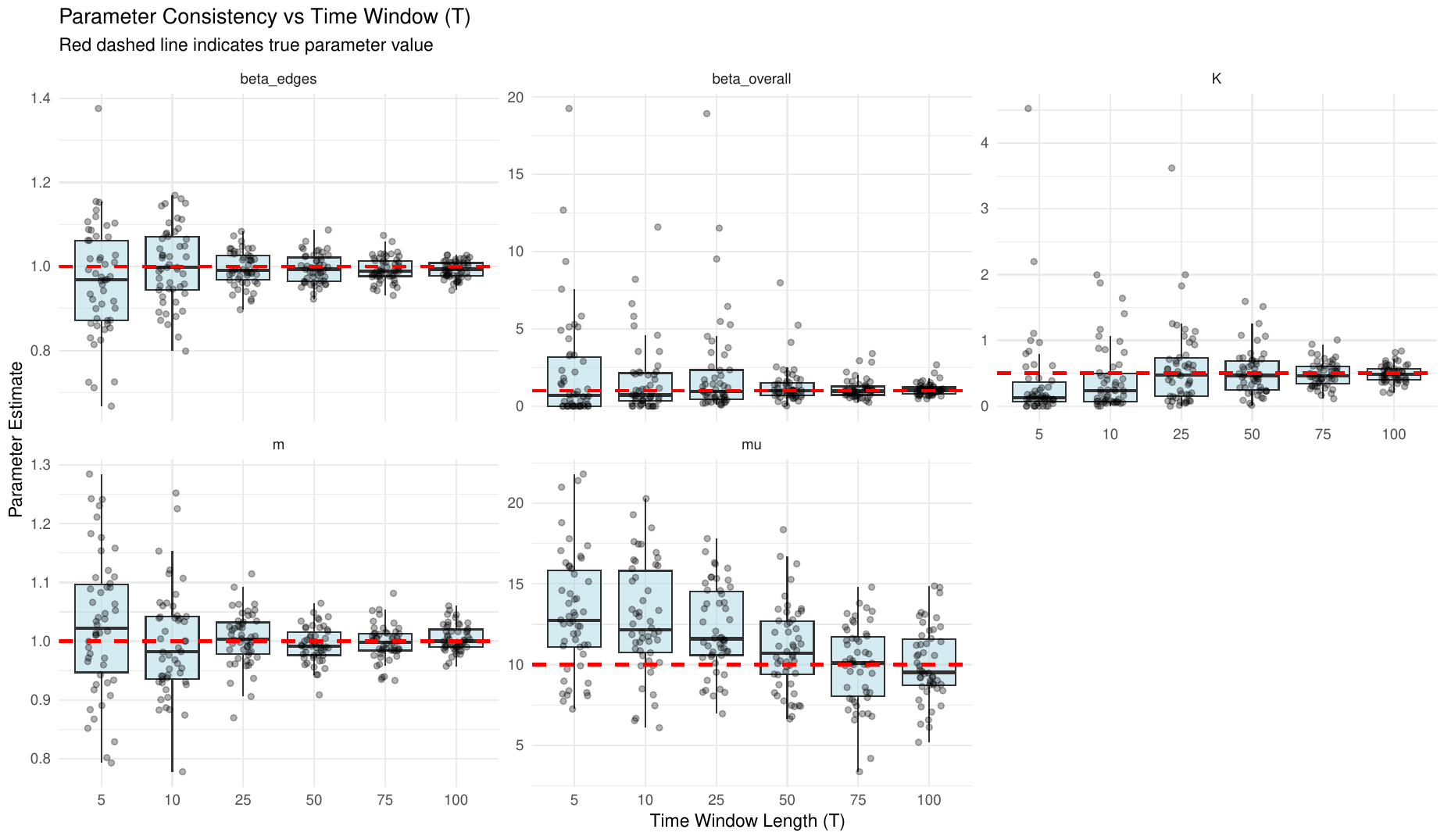}
    \caption{BA HawkesNet: Distribution of parameter estimates by time window $T$. Red dashed lines indicate the true parameter values. The tightening of the boxplots around the truth demonstrates the decreasing variance of the MLE as data increases.}
    \label{fig:BA_consistency_boxplots}
\end{figure}

\begin{figure}[H]
    \centering
    \includegraphics[width=1.0\textwidth]{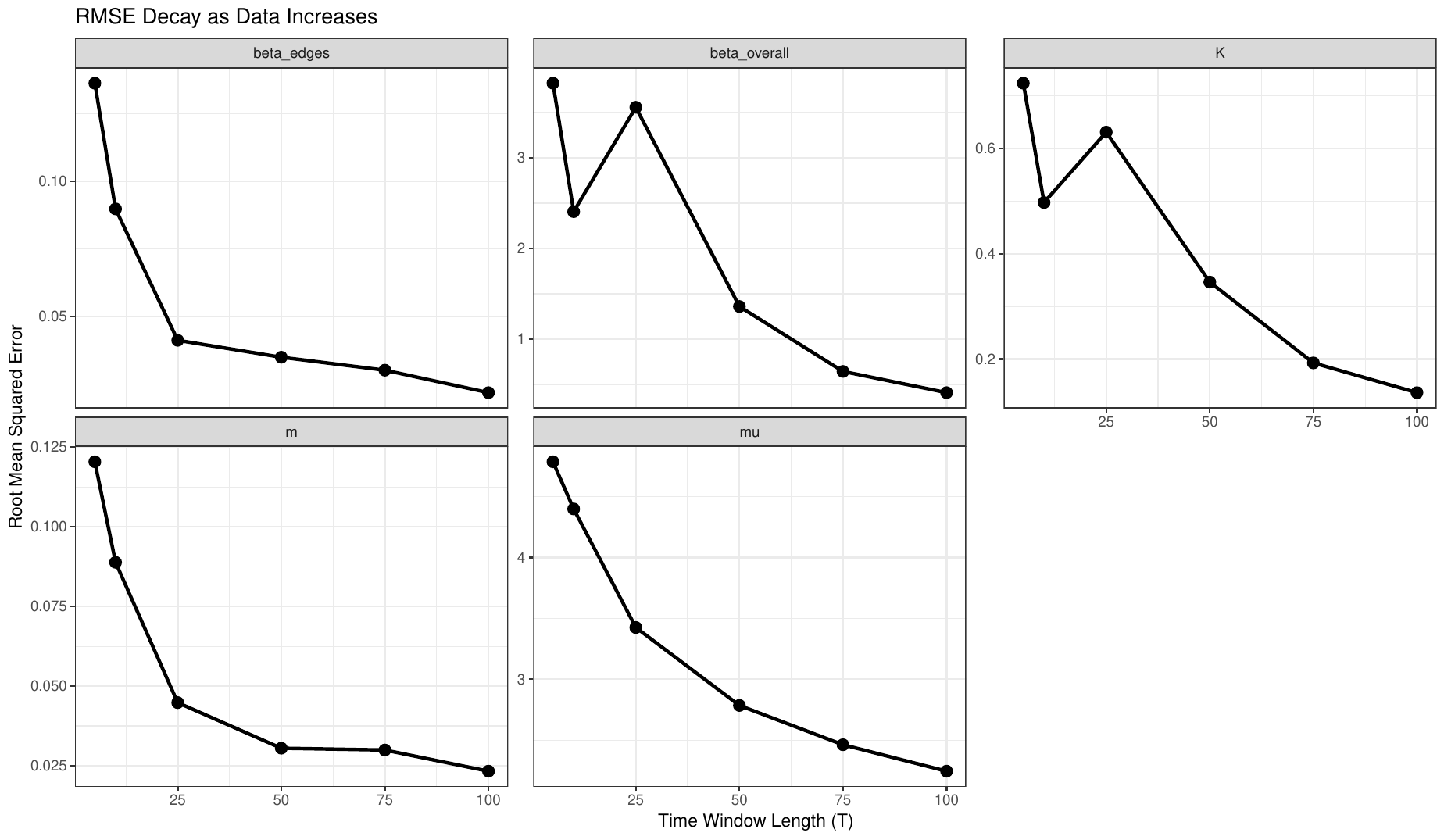}
    \caption{BA HawkesNet: RMSE versus time window $T$ by parameter. The monotonic decay in RMSE for all parameters confirms the consistency of the estimation procedure.}
    \label{fig:BA_consistency_rmse}
\end{figure}

We considered windows $T \in \{5, 10, 25, 50, 75, 100\}$. The BA HawkesNet results demonstrate clear convergence across all parameters ($\lambda_{\emptyset}, \beta,\tau, m$). Figure~\ref{fig:BA_consistency_boxplots} shows the contraction of the sampling distribution, while Figure~\ref{fig:BA_consistency_rmse} illustrates the corresponding decay in RMSE.

\subsubsection{CS Model Consistency.} 

\begin{figure}[H]
    \centering
    \includegraphics[width=1.0\textwidth]{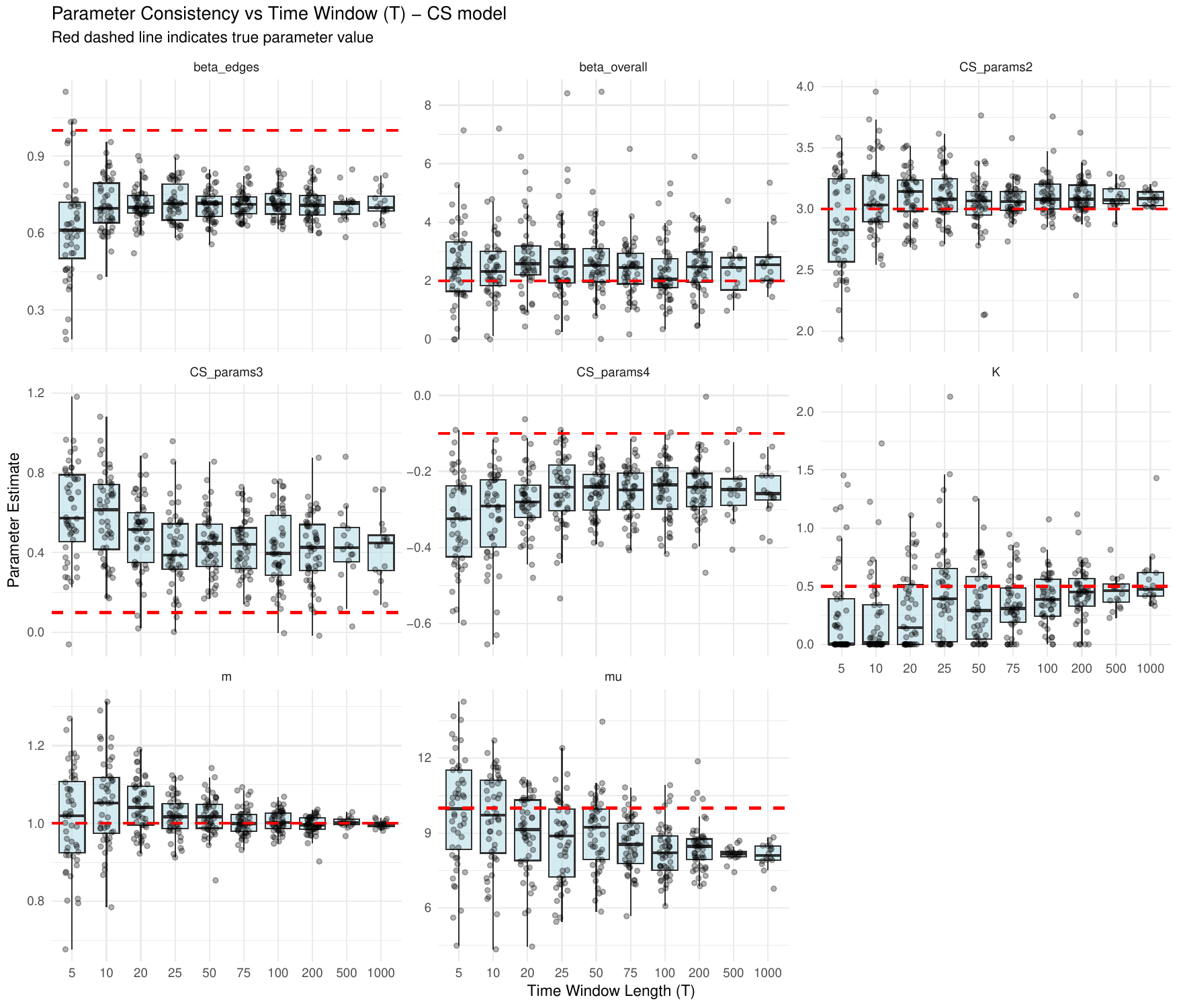}
    \caption{CS HawkesNet: Distribution of parameter estimates by time window $T$. Even with the higher-dimensional structural parameter space, the estimators center on the true values as the observation window lengthens.}
    \label{fig:CS_consistency_boxplots}
\end{figure}

The Change Statistic (CS) model involves a higher-dimensional parameter space, including structural terms (edges, triangles, $k$-stars) and nodal growth parameters ($\lambda_{\text{nodes}}$). Figure~\ref{fig:CS_consistency_boxplots} shows the sampling distributions for the CS parameters over $$T \in \{5, 10, 25, 50, 75, 100,200,500,1000\}.$$  Despite the increased complexity of the mark distribution, the model exhibits similar convergence properties.

We note that the CS model converges to values slightly different from the true parameter values. This particularly affects the 2-star and 3-star terms, which are highly correlated. The model is able to trade one off against the other and effectively find parameters that maximize the likelihood; in other words, the likelihood function is flat in this direction. A similar issue occurs with $\lambda_{\emptyset}$ (mu in the \texttt{hawkesNet} package) and $\beta_{\text{edges}}$. Again, the likelihood function is flat along the $\lambda_{\emptyset}$ and $\beta_{\text{edges}}$ direction. However, in practical applications, this would not result in a materially different interpretation.

\subsection{Finite $T$ MLE Performance}\label{sec:mle_dist}

\begin{figure}[!htb]
    \centering
    \includegraphics[width=0.95\textwidth]{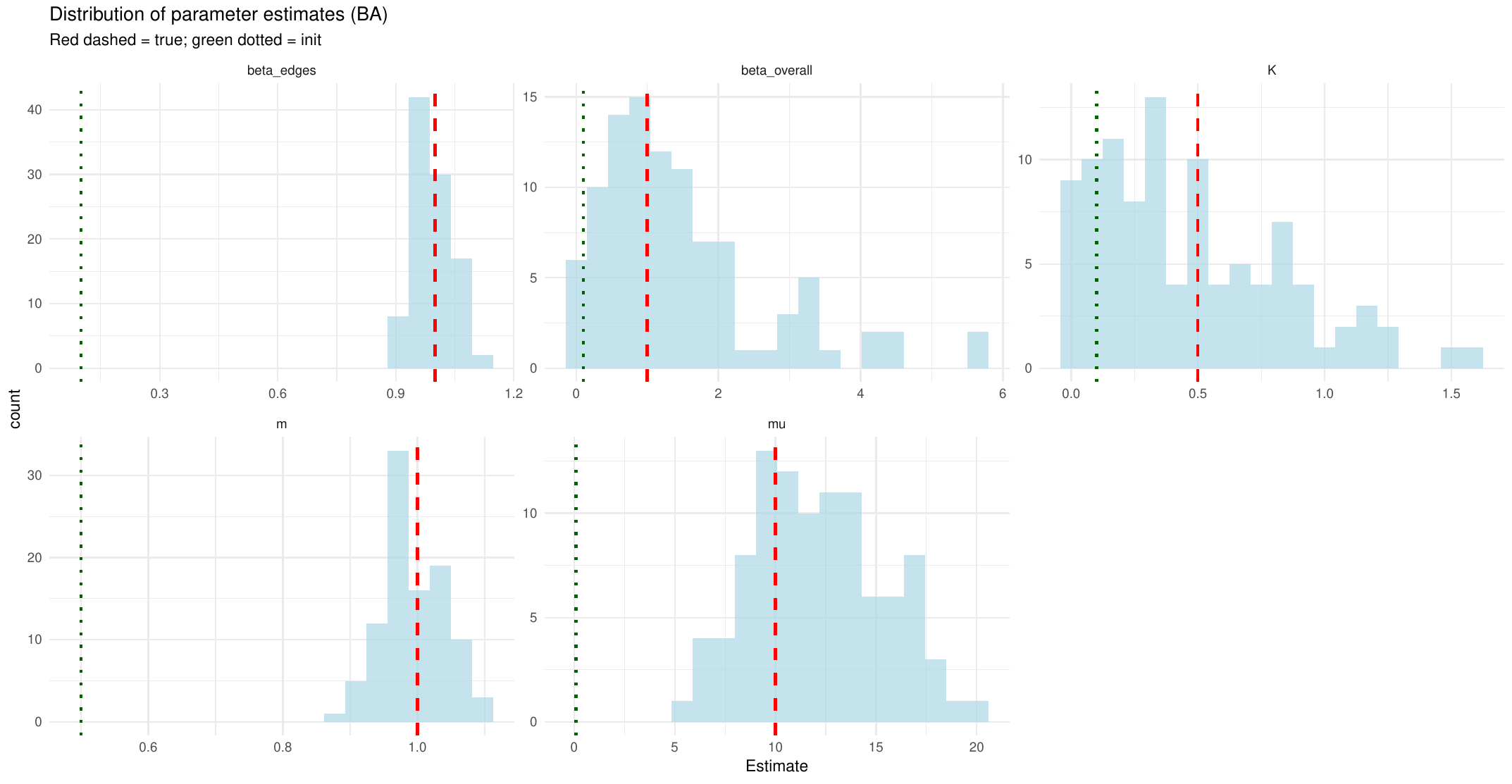}
    \caption{Histogram of the simulate MLEs from the BA simulation study}
    \label{fig:BA_sim_study_shape}
\end{figure}

Figure \ref{fig:BA_sim_study_shape} shows the distribution of the simulated parameter estimates in the simulation study for the BA hawkesNet. This demonstrates near normality even for only around $1000$ events. The Hawkes parameters seem to be typically more challenging to estimate than the mark PMF parameters.

\subsection{Supercritical vs.\ stable regime}\label{sec:app_explosivity_BA}

When the Hawkes memory decays slowly (small $\beta$) and the triggering weight $K$ is close to or above $\beta$, the integrated intensity can diverge and the process becomes supercritical (i.e. exponentially increasing): the number of events grows very quickly in time. When $\beta$ is larger and $K$ is moderate, the process is stable and event counts grow in a controlled way. To illustrate this, we compare two BA HawkesNet specifications over a short window $[0, T_{\text{expl}}]$ (e.g.\ $T_{\text{expl}} = 5$): a ``supercritical'' set with small $\beta$ and $K \approx 1$, and a ``stable'' set with larger $\beta$ and $K = 0.5$.

Table~\ref{tab:explosive_comparison_BA} reports the parameter settings alongside summary statistics from a single run of each specification.

\begin{table}[!htb]
    \centering
    \begin{tabular}{lcc}
        \toprule
        Quantity & Stable & Supercritical \\
        \midrule
        $\beta$ (Hawkes decay) & 2.0 & 0.1 \\
        $K$ (triggering weight) & 0.5 & 0.99 \\
        $\tau$ (mark decay) & 1.0 & 0.1 \\
        \midrule
        Max.\ degree in final network & 16 & 90 \\
        Total events  & 70 & 914 \\
        \midrule
    \end{tabular}
    \caption{Supercritical vs.\ stable BA HawkesNet: parameter settings and summary of one run.}
    \label{tab:explosive_comparison_BA}
\end{table}

\begin{figure}[!htb]
    \centering
    \includegraphics[width=0.85\textwidth]{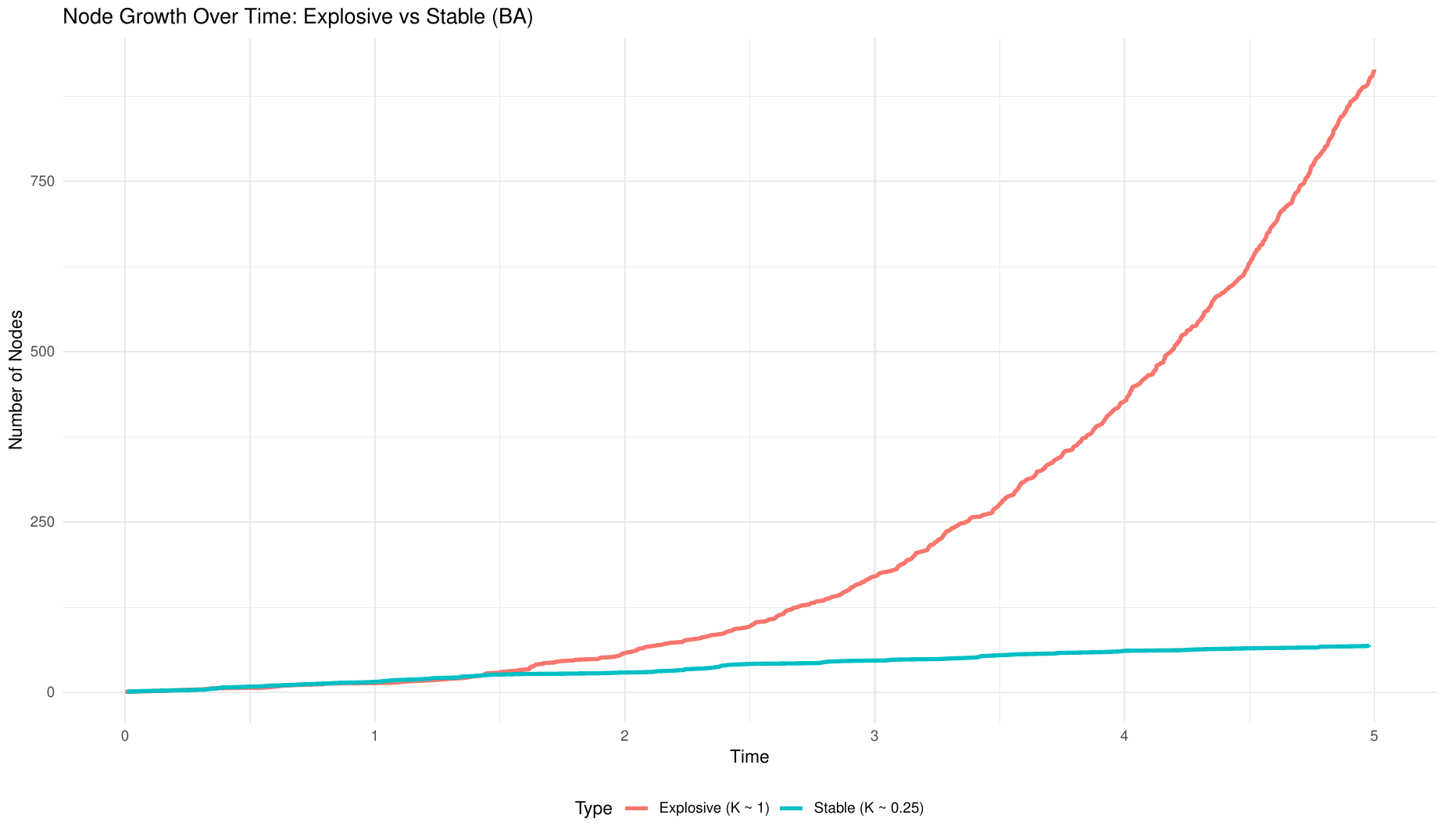}
    \caption{Cumulative number of events $N(t)$ vs.\ time $t$ for one stable and one supercritical BA HawkesNet run. Supercritical run (small $\beta$, $K \approx 1$) grows rapidly; stable run remains moderate.}
    \label{fig:BA_explosive_cumulative}
\end{figure}

Figure~\ref{fig:BA_explosive_cumulative} plots the cumulative event count $N(t)$ against time for each specification. Under the stable parameters, $N(t)$ increases roughly linearly, consistent with a stationary rate slightly above $\lambda_{\emptyset}$.

\begin{figure}[!htb]
    \centering
    \includegraphics[width=0.85\textwidth]{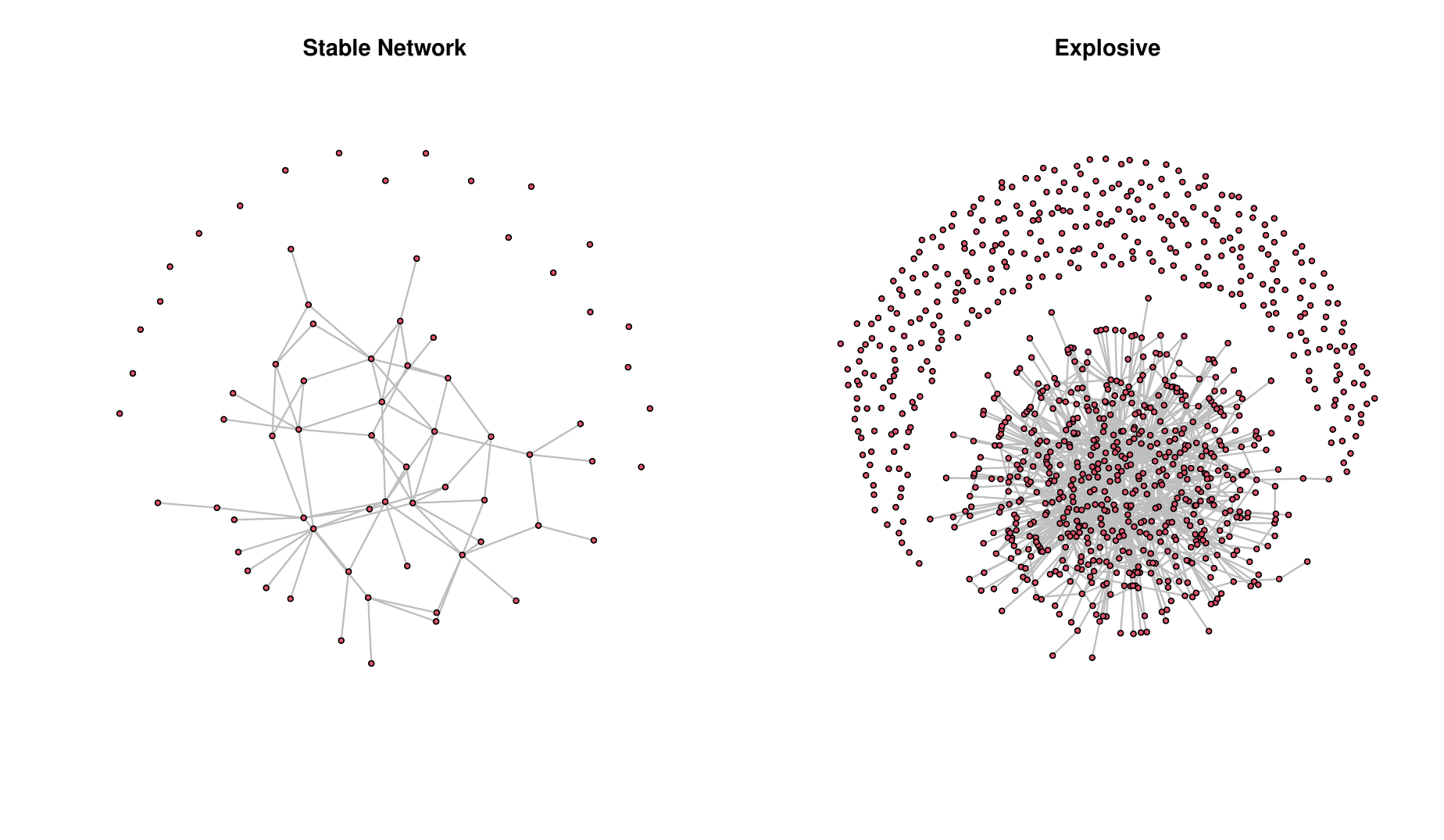}
    \caption{Network visualization of Supercritical vs stable regime in finite time window}
    \label{fig:BA_explosive_node_growth}
\end{figure}
 Under the nonstationary (supercritical) parameter settings, $N(t)$ exhibits the characteristic convex, superlinear growth of a supercritical process: each event triggers further events faster than the kernel can decay, producing an accelerating cascade. The divergence is already stark by $t = 2$--$3$, well before the end of the window. Figure~ \ref{fig:BA_explosive_node_growth} visualizes this growth of the network.
\break

\newpage

\section{Additional Application Details}

\subsection{Dataset Preprocessing and Network Construction}
\label{sec:supp_data_processing}

The raw dynamic contact network from the ACM Hypertext 2009 conference consists of $28\,468$ timestamped records of mutual proximity spanning approximately three days. Participants wore radio-frequency badges that registered face-to-face proximity at a discrete temporal resolution of 20 seconds. 

For our analysis, we restrict attention to the longest uninterrupted session (the first day, approximately 16 hours). To analyze network formation, the network is treated as undirected and only first contacts are retained, ensuring that the edge set grows monotonically. A node's arrival time is defined as the exact time of its first appearance in any recorded contact, meaning the node set is also monotonically growing.

Because the original data are recorded at a discrete 20-second resolution, many recorded contacts appear simultaneously. While the HawkesNet framework inherently accommodates simultaneous node and edge arrivals within a single mark, batching the data in this manner reduces the effective number of events to approximately 400, which can lead to poor parameter identifiability and instability during estimation. 

To emulate the underlying continuous-time nature of human interaction and strictly separate event times, we jitter the observed timestamps by adding independent Gaussian noise ($\sigma = 0.01\,\text{s}$) so that no two events share the exact same time. Finally, all event times are linearly rescaled to the unit interval $[0,1]$ to maintain numerical stability during likelihood optimization.

\subsection{Goodness of Fit}\label{app:RCT_GOF}

Assessing goodness of fit (GOF) for marked point processes is difficult, and although various \textit{ad-hoc} options exist in the literature \citep{clements2011residual,clements2012evaluation, baddeley2005residual, schoenberg2003multidimensional} none are theoretically robust. Rescaled residuals are unit Poisson distributed for a correctly specified conditional intensity function via the Random Time Change theorem \cite[cf. Proposition 7.4.VI]{daley2003introduction}. This result allows for assessment of model fit with respect to the time-dimension, and p-values can be calculated from a Kolmogorov-Smirnov (KS) test, however this does not account for the mark portion.

\subsection{Tables and Figures from \ref{sec:hypertext_gof}}

\begin{table}[htbp]
  \centering
\caption{CS HawkesNet fit to Hypertext 2009 conference interaction network.The top block shows parameters fixed for identifiability, the middle section shows Hawkes parameters and the bottom block shows mark PMF distribution parameters.}
  \label{tab:ht_fit1}
  \begin{tabular}{lrr}
    \toprule
    Parameter & Estimate & Std.\ error \\
    \midrule
    $K$ (fixed)                       & 1.000   & --- \\
    $\lambda_{\text{nodes}}$ (fixed) & 0.106   & --- \\
    \midrule
    $\lambda_{\emptyset}$                            & 844.300 & 36.542 \\
    $\beta_{\text{overall}}$         & 11.734  & 3.059 \\
    \midrule
    $\beta_{\text{edges}}$           & 27.713  & 2.281 \\
    $m$                              & 1.000   & 0.033 \\
    edges                            & $-8.285$ & 0.089 \\
    gwdegree.0.5                     & 3.458   & 0.128 \\
    gwesp.0.5                        & 0.352   & 0.042 \\
    \bottomrule
  \end{tabular}

\end{table}

\begin{figure}[!htbp]
  \centering
  \includegraphics[width=0.85\textwidth] {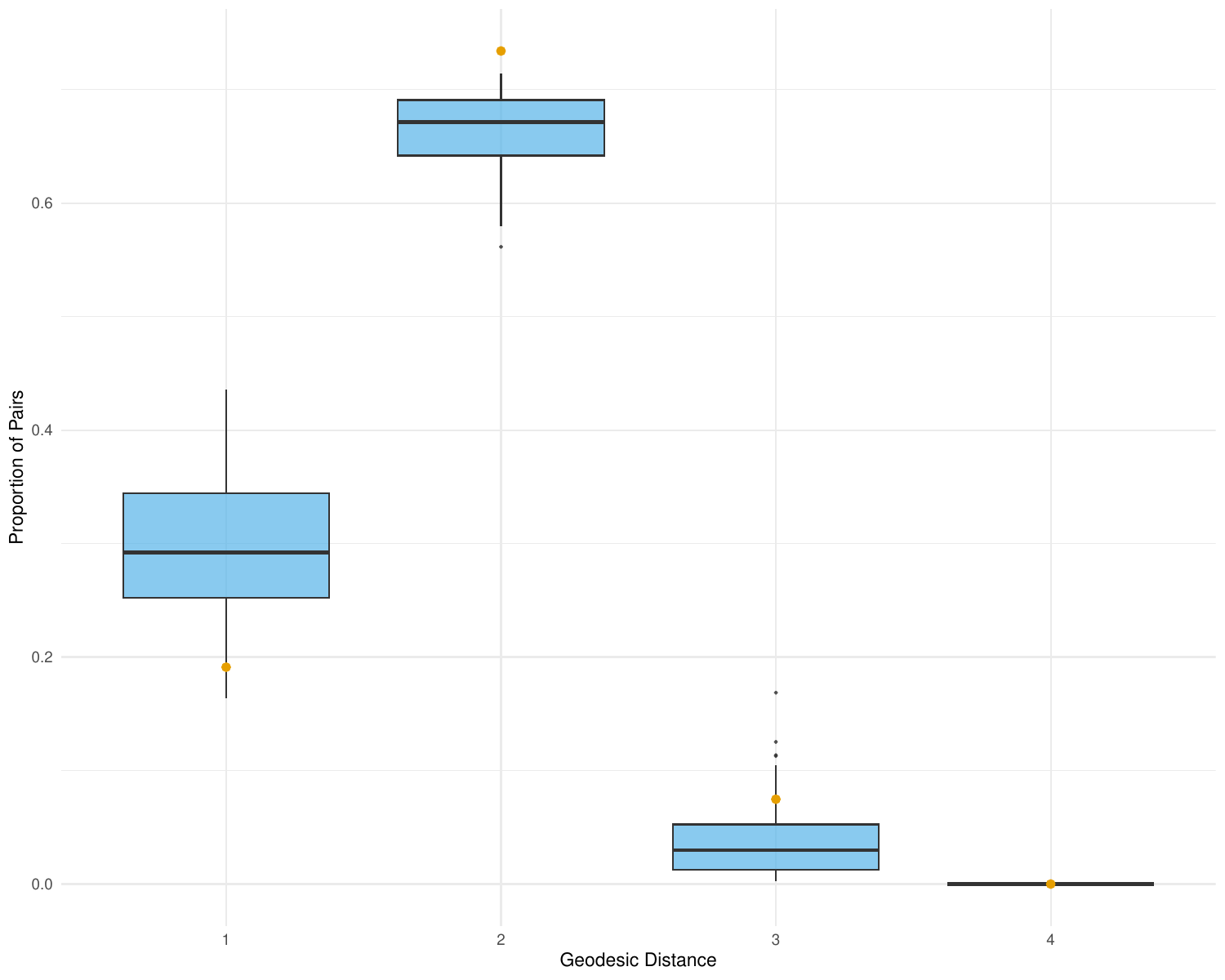}
  \caption{GOF: Geodesic distance distribution.}
  \label{fig:ht_gof_geodist}
\end{figure}

\begin{figure}[!htbp]
  \centering
  \includegraphics[width=0.9\textwidth]{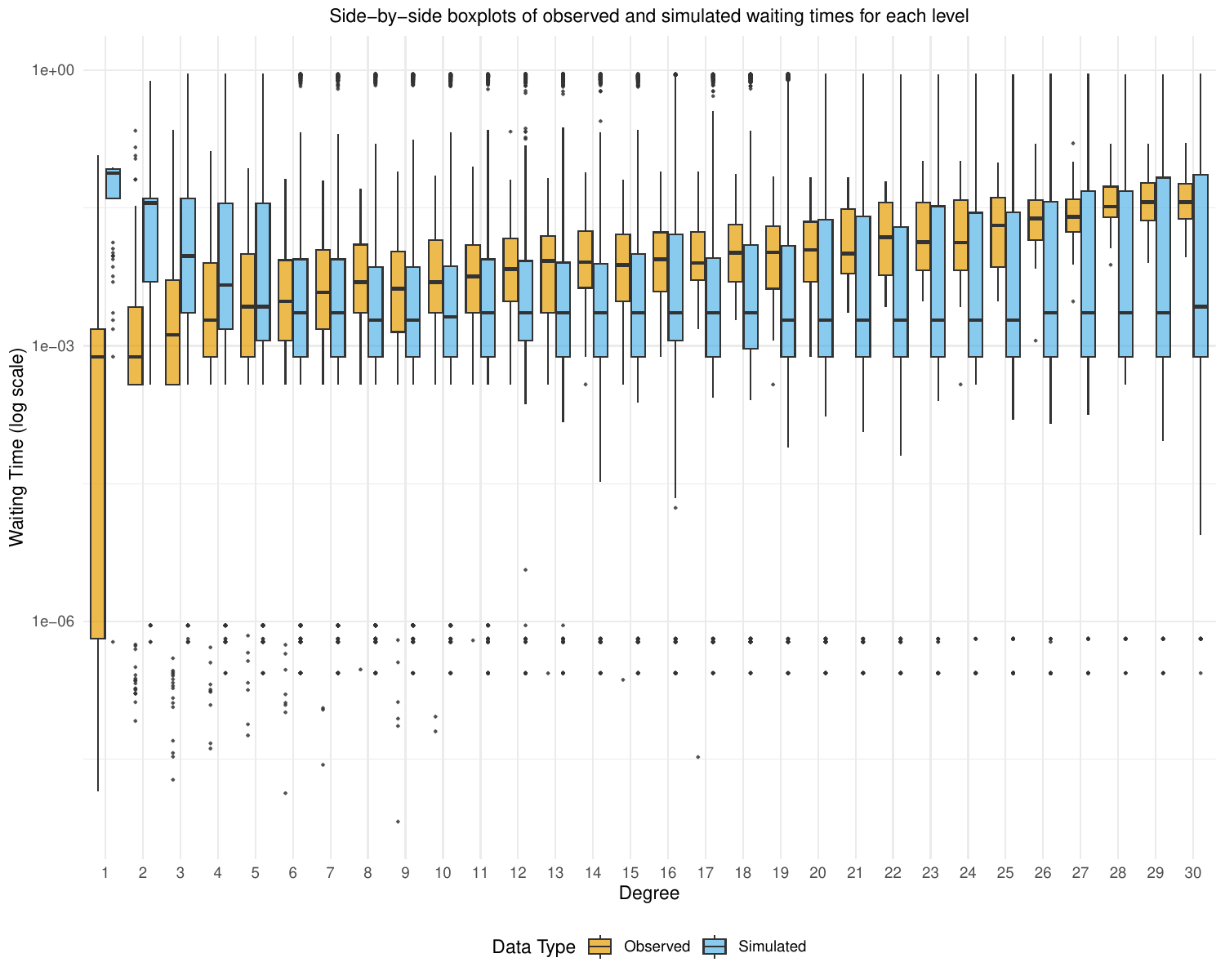}
  \caption{GOF: Waiting times between motif formation for degrees}
  \label{fig:ht_gof_waiting_degs}
\end{figure}

\begin{figure}[!htbp]
  \centering
  \includegraphics[width=0.9\textwidth]{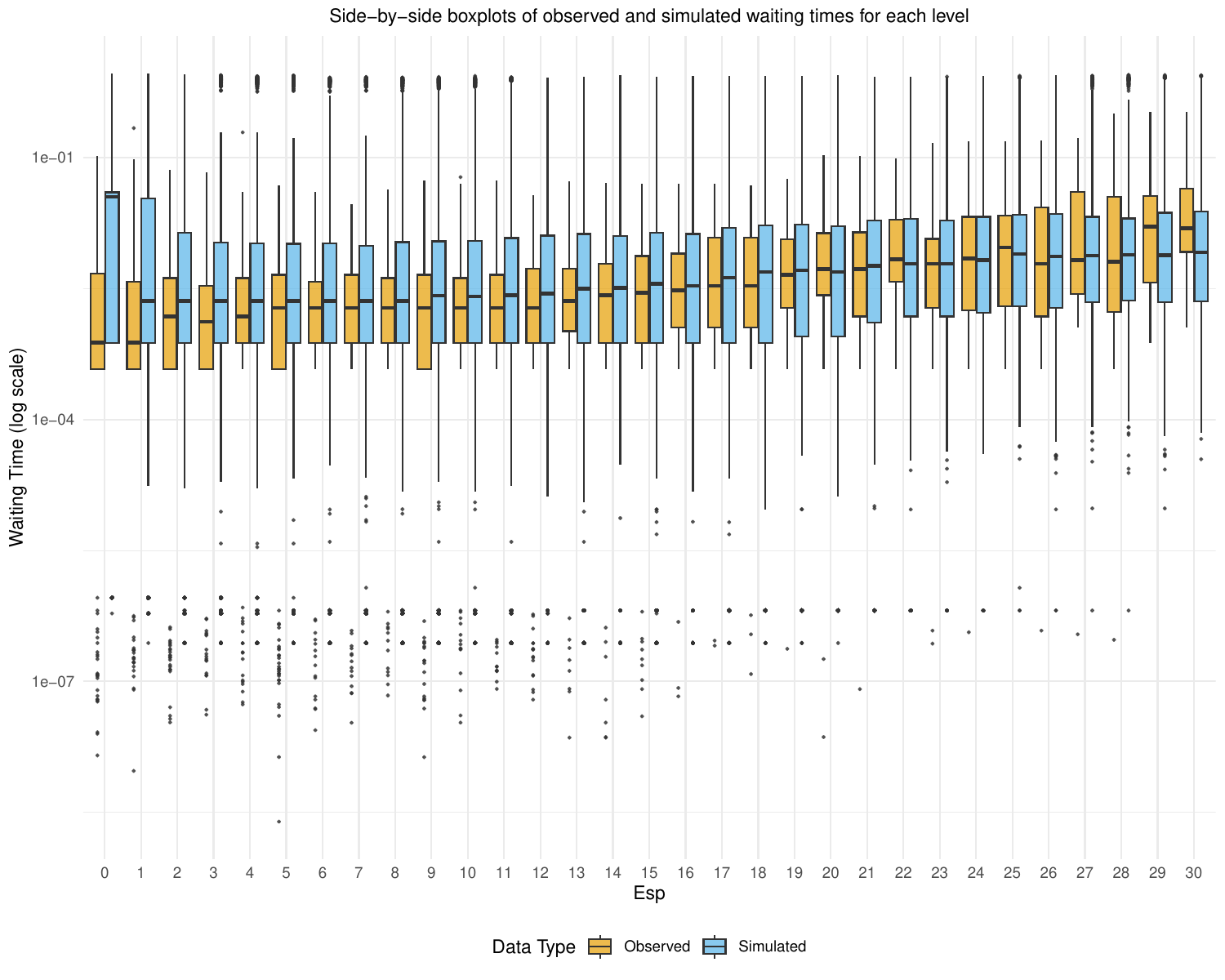}
  \caption{GOF: Waiting times between motif formation for ESP}
  \label{fig:ht_gof_waiting_esp}
\end{figure}
\break

\subsection{Reproducibility and computational time}

The raw data are from the SocioPatterns collaboration \citep{sociopatterns2025} and are bundled with the \texttt{hawkesNet} package as \texttt{ht09\_contact\_list.dat}. The fitting script and SLURM configuration are in \texttt{inst/hypertext\_conference/}. The fit and GOF with one core per GOF simulation on data took approximately $10$ minutes running on NeSI’s Mahuika cluster, utilizing an AMD EPYC 7763 (Milan) processor nodes.

The hawkesNet package is highly optimized and scalable to at least 10000 events, making it broadly available to practitioners. For example the fit in the application took around 10 minutes running on NeSI.

\subsection{Additional Model Fits}\label{app:bad_fit}

As we are able to fit triangle and star model specifications without explosion due to the regularizing effect of time decay, we initially believed this to be a parsimonious model for the Hypertext conference. Table \ref{tab:ht_fitA} shows the estimated fit for this model.

\begin{table}[htbp]
  \centering
  \caption{CS HawkesNet fit to Hypertext 2009 conference interaction network. The top block show parameters fixed for identifiability, the middle section shows Hawkes parameters and the bottom block show mark PMF distribution parameters.}
  \label{tab:ht_fitA}
  \begin{tabular}{lrr}
    \toprule
    Parameter & Estimate & Std.\ error \\
    \midrule
    $K$ (fixed)                      & 1.000   & --- \\
    $\lambda_{\text{nodes}}$ (fixed) & 0.106   & --- \\
    \midrule
    $\lambda_{\emptyset}$            & 752.754 & 33.658 \\
    $\beta_{\text{overall}}$         & 6.352   & 0.974 \\
    \midrule
    $\beta_{\text{edges}}$           & 13.869  & 1.148 \\
    $m$                              & 1.477   & 0.048 \\
    edges                            & $-4.980$ & 0.055 \\
    triangles                        & 0.131   & 0.012 \\
    star.2                           & $-0.173$ & 0.005 \\
    star.3                           & 0.005   & 0.000 \\
    \bottomrule
  \end{tabular}
\end{table}

In particular, the triangle parameter is positive while the 2-star parameter is negative, with the 3-star parameter the opposite sign, as is often observed in ERGM-style models. 

\begin{figure}[htbp]
  \centering
  \includegraphics[width=0.85\textwidth] {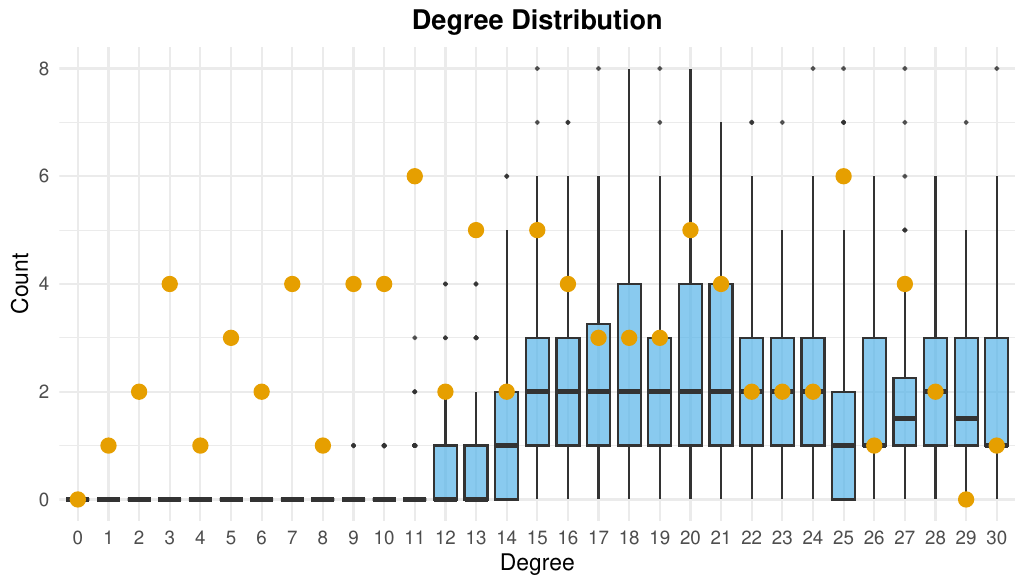}
  \caption{GOF: Degree distribution. Observed (orange points) vs.\ simulated (blue boxplots) for the triangle star model}
  \label{fig:ht_gof_degree_bad}
\end{figure}

However, the $m$ estimate is greater than $1$. Despite only one edge being added per event in the observed data, this suggests fit may be poor. Moreover, upon inspection of the degree GOF plot it becomes clear that this model does not recreate the structure of the observed network well. We believe that the strong transitive closure observed in the network pushes the MLE towards a positive triangle parameter, yet to prevent explosion the 2-star parameter is pushed towards the negative, while trying to add more edges at each step to counteract this. This produces extremely unrealistic networks in which there are very few low-degree nodes as seen in Figure \ref{fig:ht_gof_degree_bad}; conclusions from such an ill--fitting model should be treated with extreme caution.

\end{document}